\documentclass[12pt,english]{article}
\pdfoutput=1
\usepackage[T1]{fontenc}
\usepackage[utf8]{inputenc}
\usepackage{subfigure,lmodern, amsmath,amssymb, graphicx, pifont, adjustbox, bm, xcolor}
\usepackage{amsfonts}
\usepackage{geometry}
\usepackage{comment}
\usepackage{mathtools}
\usepackage{float}
\usepackage{slashed}
\usepackage{cite}
\usepackage{ragged2e}
\usepackage{etoolbox}
\usepackage{array}

\apptocmd{\thebibliography}{\justifying\setlength{\leftskip}{7.4mm}}{}{}
\makeatletter\g@addto@macro\bfseries{\boldmath}\makeatother

\usepackage{stackengine}
\usepackage{esint}
\usepackage[unicode=true,pdfusetitle,
 bookmarks=true,bookmarksnumbered=false,bookmarksopen=false,
 breaklinks=false,pdfborder={0 0 1},backref=false,colorlinks=true]
 {hyperref}
\hypersetup{pdftitle={Causality, Unitarity, and the Weak Gravity Conjecture},
 pdfauthor={Nima Arkani-Hamed, Yu-tin Huang, Jin-Yu Liu, and Grant N. Remmen},
 citecolor=black,linkcolor=black,urlcolor=black}

\newcommand{\appendixref}[1]{\hyperref[#1]{appendix~\ref{#1}}}
\def\equationautorefname~#1\null{eq.\,(#1)\null}
\usepackage{breakurl}
\usepackage[hang,flushmargin]{footmisc} 
\allowdisplaybreaks

\usepackage{changepage}

\newcommand{\eq}{\begin{equation}}
\newcommand{\eqe}{\end{equation}}
\newcommand{\eqa}{\begin{eqnarray}}
\newcommand{\eqae}{\end{eqnarray}}
\newcommand{\be}{\begin{equation}}
\newcommand{\ee}{\end{equation}}
\newcommand{\bea}{\begin{eqnarray}}
\newcommand{\eea}{\end{eqnarray}}
\newcommand{\Fig}[1]{Fig.~\ref{#1}}

\newcommand{\Eq}[1]{Eq.~(\ref{#1})}

\newcommand{\Sec}[1]{Sec.~\ref{#1}}

\newcommand{\App}[1]{App.~\ref{#1}}

\newcommand{\SB}[1]{ [#1] }					
\newcommand{\AB}[1]{ \langle #1 \rangle }	
\newcommand{\ASB}[1]{ \langle #1 ] }		

\newcolumntype{P}[1]{>{\centering\arraybackslash}p{#1}}

\begin{document}

\interfootnotelinepenalty=10000
\baselineskip=18pt
\hfill

\vspace{1cm}
\thispagestyle{empty}
\begin{center}
{\LARGE \bf
Causality, Unitarity, and \\[1mm] the Weak Gravity Conjecture}\\
\bigskip\vspace{1cm}
\begin{adjustwidth}{-0.5in}{-0.5in}\begin{center}{\large Nima Arkani-Hamed,${}^{a}$ Yu-tin Huang,${}^{b,c}$ Jin-Yu Liu,${}^{b}$ and Grant N. Remmen${}^{d,e}$}\end{center}\end{adjustwidth}
\vspace{7mm}
{
\it \small ${}^a$School of Natural Sciences, Institute for Advanced Study, Princeton, NJ 08540, USA
${}^b$Department of Physics and Astronomy, National Taiwan University, Taipei 10617, Taiwan
${}^c$Physics Division, National Center for Theoretical Sciences, Taipei 10617, Taiwan\newline
${}^d$Kavli Institute for Theoretical Physics, University of California, Santa Barbara, CA 93106, USA
${}^e$Department of Physics, University of California, Santa Barbara, CA 93106, USA}
\let\thefootnote\relax\footnote{e-mail: 
\url{arkani@ias.edu}, \url{yutinyt@gmail.com}, \url{k5438777@gmail.com}, 
\url{remmen@kitp.ucsb.edu}}
 \end{center}

\bigskip
\centerline{\large\bf Abstract}
\begin{quote} \small
We consider the shift of charge-to-mass ratio for extremal black holes in the context of effective field theory, motivated by the Weak Gravity Conjecture. We constrain extremality corrections in different regimes subject to unitarity and causality constraints. In the asymptotic IR, we demonstrate that for any supersymmetric theory in flat space, and for all minimally coupled theories, logarithmic running at one loop pushes the Wilson coefficient of certain four-derivative operators to be larger at lower energies, guaranteeing the existence of sufficiently large black holes with $Q>M$. We identify two exceptional cases of nonsupersymmetric theories involving large numbers of light states and Planck-scale nonminimal couplings, in which the sign of the running is reversed, leading to black holes with negative corrections to $Q/M$ in the deep IR, but argue that these do not rule out extremal black holes as the requisite charged states for the WGC. We separately show that causality and unitarity imply that the leading threshold corrections to the effective action from integrating out massive states, in any weakly coupled theory, can be written as a sum of squares and is manifestly positive for black hole backgrounds. Quite beautifully, the shift in the extremal $Q/M$ ratio is directly proportional to the shift in the on-shell action, guaranteeing that these threshold corrections push $Q>M$ in compliance with the WGC. Our results apply for black holes with or without dilatonic coupling and charged under any number of ${\rm U}(1)$s.

\end{quote}
	
\setcounter{footnote}{0}

\newpage
\tableofcontents
\newpage

\section{Introduction}  

\begin{figure}[t]
\begin{center}
\hspace{-11mm}\includegraphics[width=0.45\columnwidth]{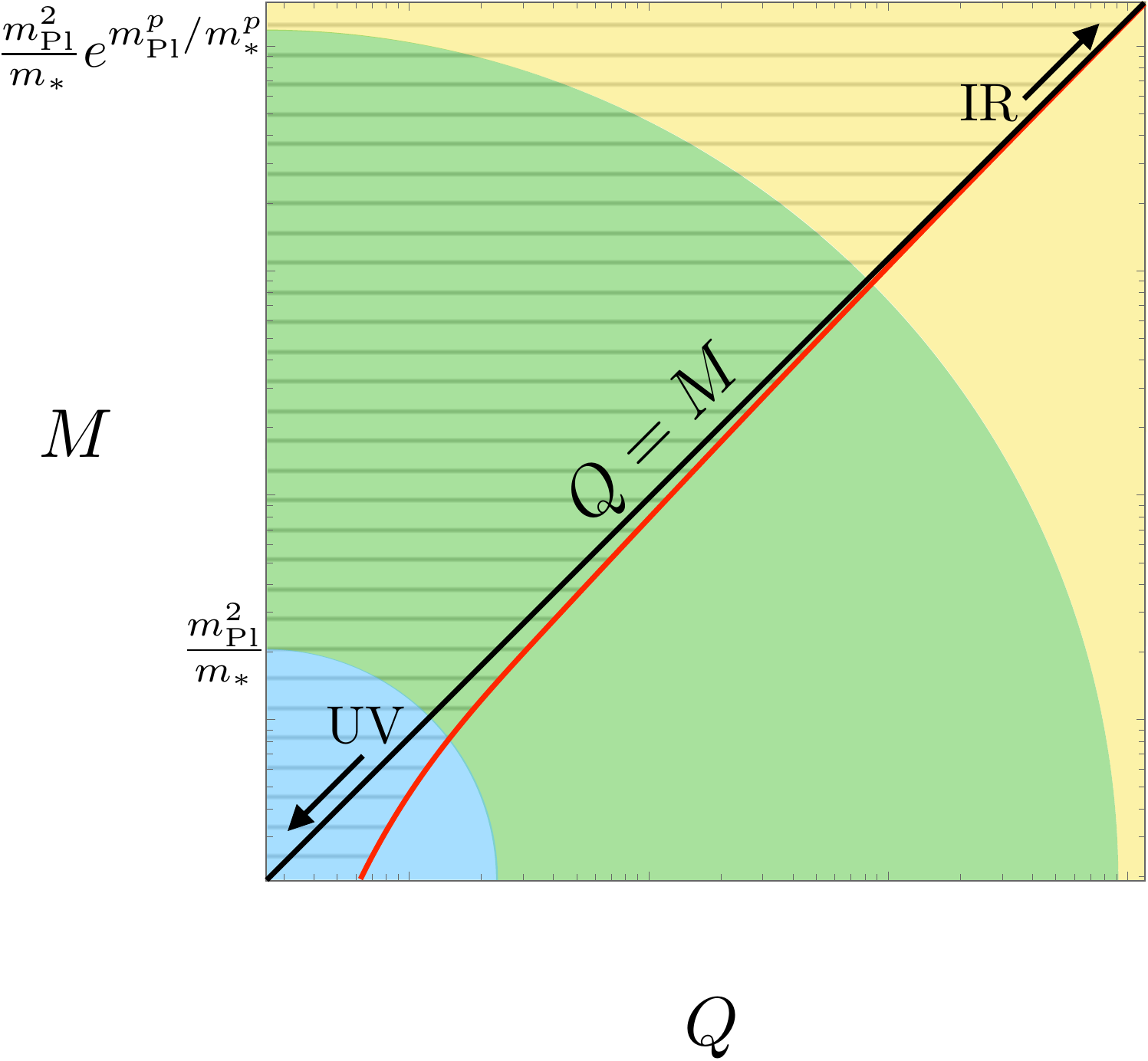}
\end{center}\vspace{-2mm}
\caption{Mass and charge parameter space for black holes (both plotted logarithmically). The unperturbed extremality line, at $Q=M$ in Planck units, is modified by higher-derivative operators to a new curve (red). We explore this curve in two regimes: the asymptotic IR (yellow), where the extremality condition is dominated by corrections induced by running of the Wilson coefficients from massless loops, and the threshold region (green), where the leading corrections are the finite coefficients induced by integrating out massive particles parametrically lighter than the Planck scale. The boundaries between these regions are stated for a completion of the higher-derivative operators at scale $m_*$. Black holes occupy the striped region, and we argue that the mass/charge curve always bends below the $M=Q$ line for healthy theories, thus satisfying the WGC.
}
\label{fig:qmcurve}
\end{figure}

The question of identifying which effective field theories (EFTs) have UV completions, subject to general principles of unitarity and causality, has long been intimately tied to our understanding of constraints associated with the consistency of quantum gravity and the the swampland program~\cite{Vafa:2005ui,Ooguri:2006in,ArkaniHamed:2006dz}. 
The most well-studied aspect of this program has been the Weak Gravity Conjecture (WGC)~\cite{ArkaniHamed:2006dz}, which states that any ${\rm U}(1)$ gauge theory coupled to gravity must be accompanied by states in the spectrum that have charge-to-mass ratio greater than unity in natural units (where ``1'' corresponds to the ratio for large extremal black holes). 
The WGC can be motivated by the demand that all black holes be able to decay~\cite{Susskind:1995da,Giddings:1992hh,tHooft:1993dmi,Bousso:2002ju,Banks:2006mm} and, in the standard model, is satisfied by light charged particles. On the other hand, for an Einstein-Maxwell system with no charged matter, the requisite charged states must be black holes. For example, consider 
\be 
{\cal L} = {\cal L}_{\rm EM} + \Delta {\cal L},
\ee
where ${\cal L}_{\rm EM} = R/2\kappa^2 - F^2/4$ is the Einstein-Maxwell Lagrangian and $\Delta {\cal L}$ comprises higher-derivative corrections. The extremal bound will be modified by the presence of $\Delta {\cal L}$~\cite{Kats:2006xp,Cheung:2018cwt,dyonic}, and the WGC requires that the extremal curve contain regions where the charge-to-mass ratio is larger than 1; see \Fig{fig:qmcurve}.

It has long been known that unitarity and analytic properties of scattering amplitudes place bounds on the coefficients of higher-derivative operators like $F^4$ and $(\partial\phi)^4$  in nongravitational EFTs~\cite{Adams:2006sv}. The earliest example of a direct link between general constraints from causality and unitarity and the WGC was the observation \cite{ArkaniHamed:2006dz} that the correct signs for the $F^4$ corrections also give a correct-sign shift leading to $Q>M$ for extremal black holes. Causality shows up in various guises in the context of EFT bounds, including the impossibility of reliable observation of global superluminality~\cite{Adams:2006sv} and the sub-$s^2$ scaling of the amplitude in the Regge limit~\cite{Arkani-Hamed:2020blm}.
In a gravitational EFT, forward scattering bounds on four-derivative (e.g., $R^2$) operators encounter a well-known subtlety involving the $t$-channel singularity associated with on-shell graviton exchange~\cite{Adams:2006sv,Cheung:2014ega,Bellazzini:2015cra,Bellazzini:2019xts,Hamada:2018dde}. This can be circumvented by considering finite-$t$ (impact parameter) dispersion relations~\cite{Caron-Huot:2021rmr}, but leading to negative lower bounds. Note, however, that in any theory of quantum gravity with a weak coupling---such as string theory---the scale $m_*$ suppressing higher-dimension operators is parameterically smaller than the Planck scale $m_{\rm Pl}$, and so the signs of the leading operators suppressed by $m_*$ are determined by the standard nongravitational dispersive arguments. And in theories with no separation between $m_*$ and $m_{\rm Pl}$, the ``higher-dimension operators'' are in fact dominated by logarithmic running generated by massless loops in the low-energy theory, and so it is these logarithmic running contributions that must be studied.

In this paper, we will investigate what causality and unitarity can tell us about the $Q/M$ curve in generality. We begin with the asymptotic IR, where as mentioned above the leading corrections to the effective action are dominated by logarithmic running. This asymptotic IR region is relevant for black holes with radii that are exponentially larger than the cutoff scale, with $r_{\rm H} \sim m_*^{-1} \exp(m_{\rm Pl}^p/m_*^p)$ for some power $p$, where the large logs can eventually supercede the presence of any higher-dimension operators suppressed by $m_*$. Note that this depends on having $D=4$ spacetime dimensions, to which we restrict our analysis; in higher dimensions, the logarithmic running generates still higher-dimension operators that can never compete with the leading $m_*$-suppressed terms. We next consider the leading threshold corrections generated by integrating out massive particles at $m_*$, whose signs can be reliably controlled when $m_*$ is parametrically smaller than $m_{\rm Pl}$. Our threshold results apply at loop level, for any number of photon species, and at arbitrary order in derivatives, while our results on the beta function allow us to sidestep complications associated with $t$-channel singularities and cubic terms like $R_{\mu\nu\rho\sigma}F^{\mu\nu}F^{\rho\sigma}$ to uncover universal behavior of asymptotically-large black holes.
The leading higher-dimension operators containing four derivatives have been investigated previously, and one can argue that the $Q/M$ curve indeed bends upward from these terms, under certain assumptions, as a consequence of consistency of black hole entropy for tree-level completions~\cite{Cheung:2018cwt,dyonic} or unitarity of single-${\rm U}(1)$ theories~\cite{Bellazzini:2019xts,Hamada:2018dde}.

In considering the asymptotic IR in \Sec{sec:bubble}, we start with pure Einstein-Maxwell theory with additional minimally coupled massless fields  and demonstrate that the beta function for the $F^4$ term (which appears in the form of the square of the stress tensor) always causes the Wilson coefficient to grow larger at low energies, pushing the $Q/M$ curve up and satisfying the WGC.
However, this cannot be the end of the story, since in extended SUSY, the $F^4$ term is protected by nonrenormalization theorems. This implies that there must be some negative contributions for the cancellation. To identify the source of negativity we turn to supergravity theories. For general $\mathcal{N}=1$ theories, the beta function generates a strictly positive $F^4$ term. For $\mathcal{N}=2$, the fact that the beta function of the graviphoton $F^4$ operator vanishes, for an arbitrary number of vector multiplets and hypermultiplets, can be traced to the negative contributions arising from the dimension-five operators $\bar{\psi}F\psi$ and $\phi F^2$. This implies that by breaking SUSY and introducing an arbitrarily large number of such operators, we can eventually generate negative Wilson coefficients. We verify that this is indeed the case for $N>137$ fermions  or $N>46$ scalars, nonminimally coupled via  the $\bar{\psi}F\psi$ of $\phi F^2$ operators, respectively. Thus in the deep IR the presence of nonminimal couplings can drive the running of $F^4$ operators to negative values.

\begin{figure}[t]
\begin{center}
\hspace{-11mm}\includegraphics[width=0.45\columnwidth]{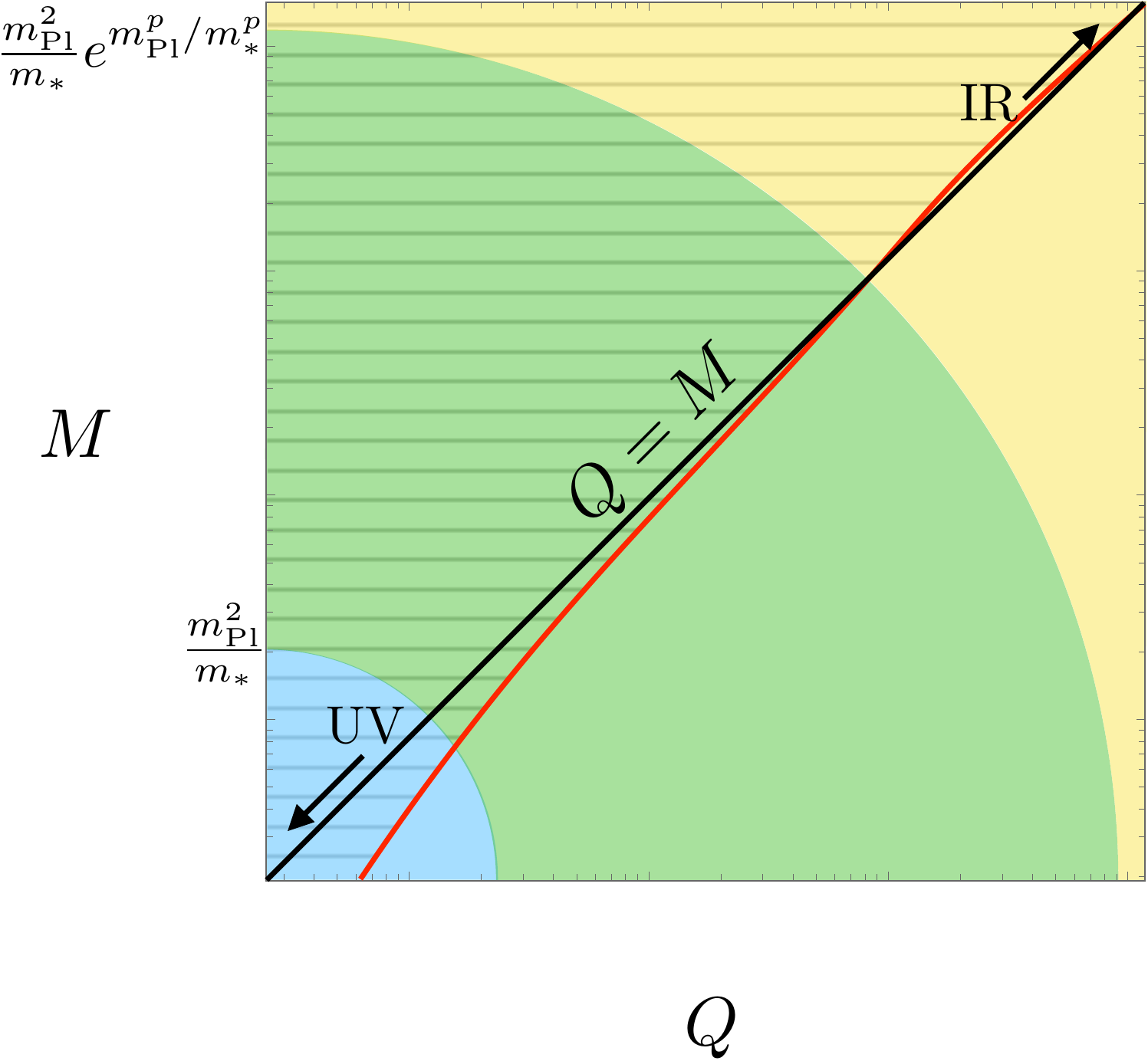}
\end{center}\vspace{-3mm}
\caption{In rare examples of nonsupersymmetric theories with a large number of species and nonminimal couplings in a specific Planckian range, the beta function  drives the running of the $F^4$ Wilson coefficient to negative values in the exponentially deep IR (yellow).
In that region, the $Q/M$ ratio for extremal black holes acquires a negative correction.
Nonetheless, we expect that in healthy theories in the threshold region (green), the finite corrections associated with sub-Planckian states in the UV completion nonetheless yield a net positive contribution to $Q/M$, so that all extremal black holes---including those with exponentially large masses---are able to decay, preserving the WGC.
}
\label{fig:qmcurvecounter}
\end{figure}

The existence of negative coefficients for $F^4$ terms raises the spectre of superluminality, but in \Sec{sec:causality} we show that this is not the case for $F^4$ terms suppressed by the Planck scale:  the time advance generated by the $F^4$ operator cannot override the gravitational time delay without exiting the validity of the EFT approximation.
We also argue that these examples do not ultimately lead to a violation of the WGC, by examining issues of UV completion and tuning in \Sec{sec:tuning}, noting that the WGC only stipulates that there exists a state with charge-to-mass ratio larger than 1 somewhere along the entire curve as we move to smaller masses. It does not require the curve to move monotonically away from 1. As we will see, once we enter the threshold region, where the shift is generated by integrating out massive states below $m_{\rm Pl}$, unitarity will tend to enforce positivity of these corrections; see \Fig{fig:qmcurvecounter}.

As we are considering threshold contributions from states at $m_{*}\ll m_{\rm Pl}$, gravitational effects are irrelevant. We will utilize the remarkable fact, reviewed in \Sec{sec:action}, that the shift of the extremal charge-to-mass ratio is in fact proportional to the value of $\Delta \mathcal{L}$ itself, evaluated on the on-shell extremal solution in the two-derivative theory~\cite{Cheung:2018cwt,dyonic}. For EFTs that do not induce nonminimal three-particle amplitudes---e.g., a theory where $R_{\mu\nu\rho\sigma}F^{\mu\nu}F^{\rho\sigma}$ can be ignored---we demonstrate in \Sec{sec:squares} that unitarity, via the generalized optical theorem, implies that  $\Delta \mathcal{L}$ can be written as a sum over squares for black hole backgrounds and is hence manifestly positive.\footnote{Past works have found that in healthy theories where such three-point couplings are present, $Q/M$ is also shifted in the correct direction~\cite{Cheung:2018cwt,dyonic,Kats:2006xp,Hamada:2018dde,Cano:2019oma,Cano:2019ycn}.} Note that this statement goes beyond Einstein-Maxwell theory and operators at leading order in derivatives, which we demonstrate using quartic Riemann corrections to Reissner-Nordstr\"om (RN) black holes and EFT deformations of dilatonic (GHS) black holes.
Remarkably, we find that the action/extremality relationship continues to hold for GHS black holes.
We conclude and discuss future directions in \Sec{sec:outlook}.

\section{Beta functions and bubble cuts}\label{sec:bubble}
We begin with the leading higher-derivative corrections to Einstein-Maxwell theory:\footnote{Via the Bianchi identity,  operators of the form $DFDF$ can be traded for the Riemann/Ricci tensor contracted with $F^2$~\cite{Deser:1974cz}.}
\be
\Delta {\cal L} = a_1 (F_{\mu\nu} F^{\mu\nu})^2 + a_2 (F_{\mu\nu} \widetilde F^{\mu\nu})^2 + b F_{\mu\nu} F_{\rho\sigma} R^{\mu\nu\rho\sigma},
\ee
where operators involving the Ricci tensor can be removed via field redefinition and the Riemann-squared operator can be dropped in four dimensions since the Gauss-Bonnet term is a total derivative. Throughout, we define $\widetilde F_{\mu\nu} = \epsilon_{\mu\nu\rho\sigma}F^{\rho\sigma}/2$ and, unless otherwise noted, all higher-derivative couplings will be normalized with appropriate powers of $\kappa^2$ to be dimensionless.

We will first consider the logarithmic running of these coefficients due to massless loops. 
That is, we will take $a_1, a_2, b$ to be dominated by large logarithmic terms, and will postpone consideration of the finite pieces that are generated by massive states in the UV to Secs.~\ref{sec:action} and \ref{sec:squares}. The beta function for each operator can be extracted from the UV divergence of the four-photon amplitude, where $a_1,a_2$ are linearly mapped to the all-plus (minus) and the two-plus, two-minus (MHV) helicity sector, and $b$ is mapped to the single-minus (plus) helicity sector. As unitarity cuts control the coefficients of the logarithms, we can easily deduce some general results for $(a_1, a_2, b)$. First, the absence of two-particle cuts for the single-helicity configuration leads to the absence of $R_{\mu\nu\rho\sigma}F^{\mu\nu}F^{\rho\sigma}$ corrections at one loop, and hence $b=0$. For the all-plus helicity configuration, the only nonvanishing two-particle cut occurs when the dimension-five operator $\phi F^2$ is present, where $\phi$ is a massless scalar. This leads to the following cut diagram:
$$\includegraphics[scale=0.5]{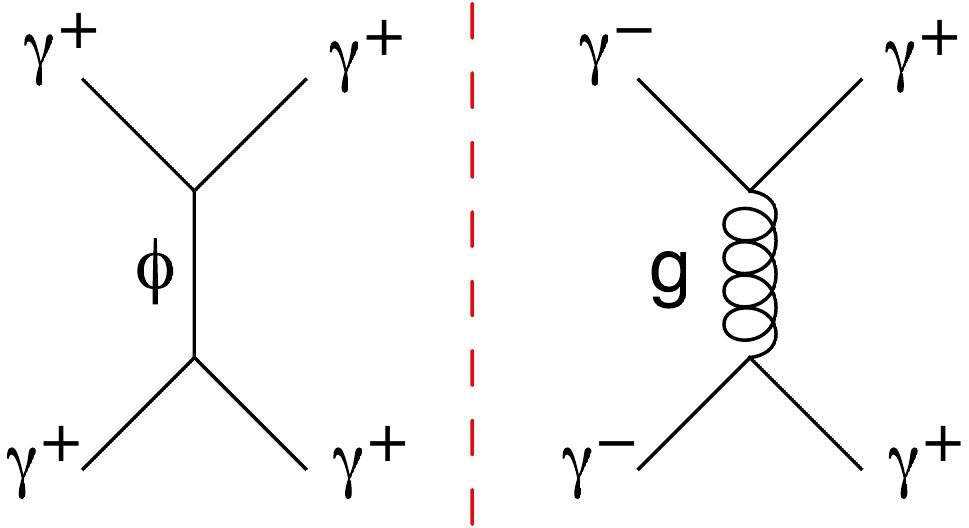}\,.$$    
Thus, in absence of a $\phi F^2$ coupling, the large logarithms will be proportional to the square of the photon stress-energy tensor,
\eq
T_{\mu\nu}T^{\mu\nu} =\frac{1}{4}(F^2)^2+\frac{1}{4}(F\widetilde{F})^2,
\eqe
which is the unique combination of the two $F^4$ operators that gives vanishing all-plus amplitude for the photon. The effective action that captures the logarithmic running therefore takes the form:
\eqa\label{Stress}
\mathcal{L}&=&\mathcal{L}_{\rm EM} + a (T_{\mu\nu})^2.
\eqae
As previously mentioned, the charge-to-mass ratio of the extremal RN black hole is modified under the presence of $ a_1 (FF)^2 + a_2 (F \widetilde F)^2$~\cite{ArkaniHamed:2006dz,Kats:2006xp,Cheung:2018cwt}, with
\eq
\left.\frac{\sqrt{Q^2 + P^2}}{\sqrt{2}M}\right|_{\rm ext} =1+\frac{16\left[a_1(Q^2 - P^2)^2 + 4a_2Q^2P^2\right]}{5(Q^2 + P^2)^3},\label{eq:shiftRN}
\eqe
where $(Q,P)$ are the electric and magnetic charges of the black hole,  respectively (with $\kappa^2$ normalized to $1$). For Eq.~\eqref{Stress}, the extremality shift goes like $4a_1=4a_2=a$.

In four dimensions, the one-loop amplitude can be cast into a scalar integral basis involving bubbles, triangles, and boxes~\cite{tHooft:1978jhc, Bern:1992em, Bern:1993kr}. As only the bubble integrals are UV-divergent, the logarithm coefficients $c_s, c_{t},c_{u}$ can be identified with the coefficients of the bubble integrals in the $s$, $t$, and $u$ channels respectively. Indeed, for an $s$-channel scalar bubble integral, we have:
\eq
c_s \int \frac{{\rm d}\ell^{4 - 2\epsilon}}{(2\pi)^4}\frac{1}{\ell^2(\ell - p_1 - p_2)^2} =c_s \left(\frac{1}{\epsilon} - \log \frac{s}{\mu^2}\right)  + \mathcal{O}(\epsilon^0).
\eqe
Recalling the dependence of $t$ and $u$ on $s$, the running of the coefficient $a$ is thus given by:\footnote{Here we only consider bubbles with two-particle cuts, since there are no UV divergences associated with bubbles on the external legs. This can be inferred from the absence of their IR counterpart. Indeed, for gravitational theories the collinear divergence cancels, and the IR divergence is universal and proportional to $\log s, \log t, \log u$~\cite{Dunbar:1995ed}. Thus, there are no IR divergences associated with massless bubbles, indicating that their UV divergences cancel as well.}
\be 
\begin{aligned}
a=-c_s\log \frac{s}{\mu^2} - c_t \log \frac{t}{\mu^2} - c_u \log \frac{u}{\mu^2}\;\xRightarrow[\;]{s\ll \mu^2}\;-(c_s + c_{t} + c_{u})\log \frac{s}{\mu^2}.\label{eq:logs}
\end{aligned}
\ee
Thus, in the deep IR where $-\log (s/\mu^2)\gg 1$, the sign of $a$ will be determined by that of  $(c_s + c_{t} + c_{u})$.

\begin{figure}
\centering
\subfigure[$c_s=-\frac{31}{30}-\frac{s^3}{4u^3} -\frac{9s^2}{8u^2}-\frac{47s}{24u}$]{\includegraphics[width=50mm]{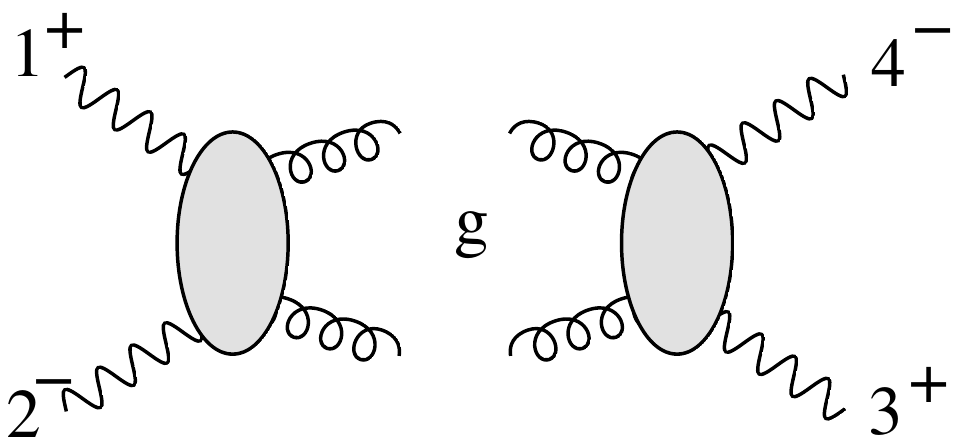}}
\quad\subfigure[$c_t=\frac{1}{20} + \frac{s^3}{4u^3}-\frac{3s^2}{8u^2} + \frac{11s}{24u}$]{\includegraphics[width=50mm]{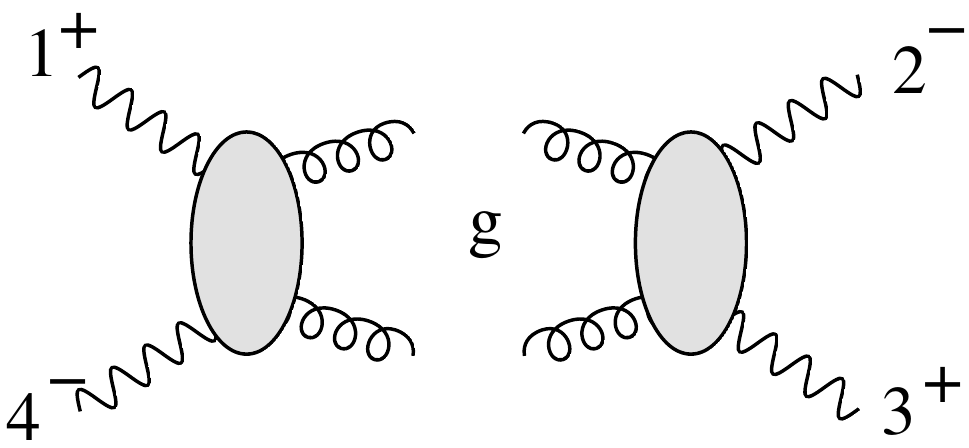}}\\
\subfigure[$c_t=\frac{87}{40} + \frac{s^3}{4u^3} + \frac{9s^2}{8u^2}+\frac{47s}{24u}$]{\includegraphics[width=50mm]{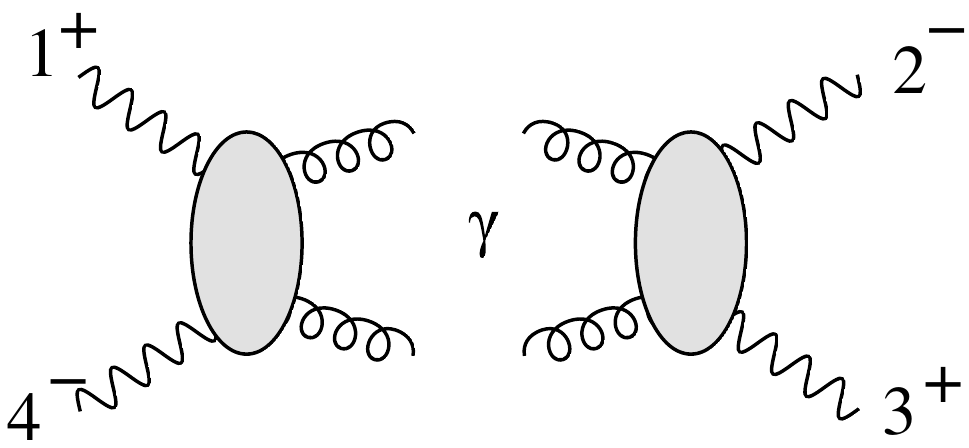}}
\quad\subfigure[$c_s=\frac{131}{120}-\frac{s^3}{4u^3}+\frac{3s^2}{8u^2}-\frac{11s}{24u}$]{\includegraphics[width=50mm]{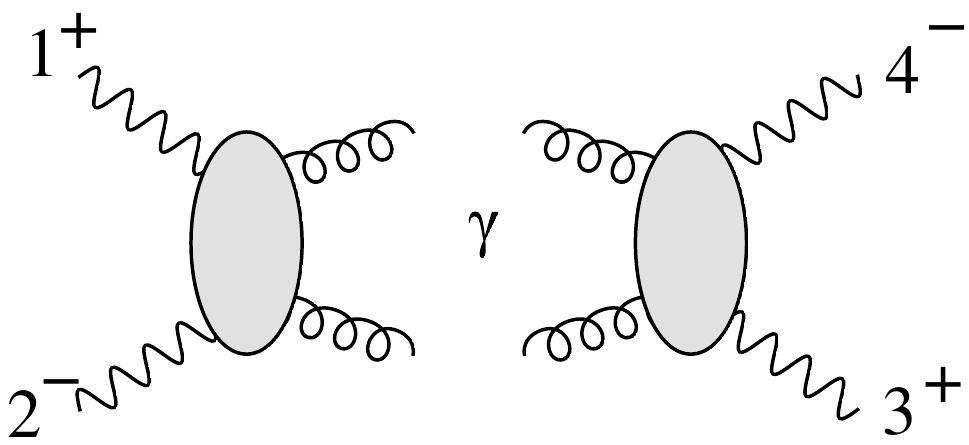}}
\caption{The $s$- and $t$-channel bubble coefficients for the graviton and photons in Einstein-Maxwell theory multiplying $(T_{\mu\nu})^2$.}\label{fig1}
\end{figure}

Let us consider the effect of minimally coupled massless fields with spin $\leq 1$. The coefficients are computed using unitarity methods devised by Forde~\cite{Forde:2007mi}, with further modifications in Refs.~\cite{ArkaniHamed:2008gz, Britto:2005ha}; see \App{app:bubble} for details. Without loss of generality, we consider the four-photon amplitude with helicities $(1^+,2^-,3^+,4^-)$. Note that for minimally coupled theories, the $u$-channel cut  $c_u$ vanishes, as on either side of the cut one has same-helicity photons. In general, each coefficient will be nonlocal, and it is only for the combination for each irreducible subset that we find a local result. For example, the coefficients for the graviton and photon in each channel are listed in \Fig{fig1}, which combine to give $137/60$. This is the well-known UV divergence of Einstein-Maxwell theory~\cite{Deser:1974cz}, and as first pointed out in Ref.~\cite{Cheung:2014ega}, its positive sign comports with the WGC in the asymptotic IR. With additional matter, the contribution for each spin is given as: 
\eq
c_s + c_t:\;\;\; \text{scalar:}\;\frac{1}{120},\;\;\; \text{fermion:}\;\frac{1}{40},\;\;\; \text{vector:}\; \frac{1}{10}\,,
\eqe
where the vectors here correspond to those that are distinct from the external one(s) under which the black hole is charged. Each indeed contributes positively.

While we have seen that adding an arbitrary number of minimally coupled matter fields only pushes the charge-to-mass ratio up, this cannot be the whole story. This is because we know that at some point there must be negative contributions, in accordance with various nonrenormalization theorems in extended supergravity theories. Understanding this cancellation in detail will shed light on the nature of negative contributions to the running of the four-(gravi)photon operator.

\subsection{Running in supergravity}  \label{sec:SUGRA}
Supergravity theories introduce two new features: the presence of a spin-$3/2$ particle and nonminimal couplings. As we will see, these two features are precisely the source of negative contributions to the running of $F^4$.

Let us begin with the one-loop UV divergence for $\mathcal{N} = 1$ Einstein-Maxwell supergravity. The four-photon divergence now receives extra contributions from the gravitino $\psi$ and photino $\lambda$. Note that the requirement that the gravitino and photino must come hand-in-hand for consistent factorization at tree level~\cite{McGady:2013sga} is also reflected in the fact that the bubble coefficients for the two contributions are individually nonlocal and only reduce to $(T_{\mu\nu})^2$ when combined. Separating the irreducible contributions, we find: \begin{center}
    \begin{tabular}{ | c | c | c | }
    \hline
    $\mathcal{N}=1$ & $a$ \\ \hline
    photon + graviton & $\frac{137}{60}$ \\ \hline
    gravitino + photino & $-\frac{1}{5}$ \\ \hline
              \end{tabular}
\end{center}
This sums to $25/12$, in agreement with Ref.~\cite{vanNieuwenhuizen:1976bg}.  Note that indeed the gravitino yields a negative contribution. However, it is not sufficient to overcome the Einstein-Maxwell contribution, which suggests adding more gravitinos, i.e., extended SUSY. But before doing so, let us add nonminimal couplings involving the Maxwell field, which in $\mathcal{N}=1$ language will be given by
\eq
g\int {\rm d}^4 x\,{\rm d}^2\theta\; \Phi W^\alpha W_\alpha+{\rm c.c.}
\eqe     
This includes the dimension-five couplings $\phi F^2$ and $\psi F \chi$, where $(\phi, \psi)$ are the scalar and fermion in the chiral multiplet and $\chi$ is the photino.
(Recall that conventional dipole couplings involving only matter fields are forbidden by rigid ${\cal N}=1$ SUSY.)
Assuming that there are $n$ chiral multiplets with such couplings, the contribution to the divergence is manifestly positive, 
\eq
\frac{1}{4}(ng - 1)^2  + \frac{n}{24} + \frac{11}{6}.
\eqe
With $\mathcal{N}=1$ SUSY, $F^4$ thus always runs to larger positive values in the IR.

For extended supergravity theories, there are no UV divergences for the one-loop four-graviphoton amplitude, irrespective of the number of matter multiplets. To understand this cancellation, let us study the $\mathcal{N}=2$ system, where the supergravity sector contains the graviton, two gravitinos, and a graviphoton, while the Maxwell multiplet contains a photon, two photinos, and a complex scalar, and the hypermultiplet contains four scalars and two fermions. Their contributions to the four-graviphoton bubble coefficients are as follows:
\begin{center}
\quad
    \begin{tabular}{ | c | c | c | }
    \hline
     $\mathcal{N}=2$ & $a$ \\ \hline
     graviphoton +  graviton & $\frac{137}{60}$ \\ \hline
     2 gravitino & $-\frac{137}{60}$ \\ \hline
   $n_m\times$(Maxwell)  photon+scalar& $-\frac{n_m}{20}$ \\ \hline
   $n_m\times$(Maxwell)  photino & $\frac{n_m}{20}$ \\ \hline
 $n_h\times$(hyper) scalar& $\frac{n_h}{30}$ \\ \hline
 $n_h\times$(hyper) fermion&  $-\frac{n_h}{30}$ \\ \hline
       \end{tabular}
\end{center}
Note that the supergravity multiplet, hypermultiplet, and super-Maxwell multiplet each cancel separately. It is enlightening to understand the source of cancellation within the matter multiplet. As previously noted, additional minimally coupled massless states with helicity $<3/2$ always contribute positively to the divergence. Thus, nonminimal coupling must be present to account for this cancellation. Indeed, for the Maxwell multiplet, the complex scalar and photon couple with the graviphoton via the dimension-five operator $\phi F_g F_m$, where $F_m$ and $F_g$ denote the matter and graviphoton field strengths, respectively. Similarly for the hypermultiplet, the fermions couple to the graviphoton with gravitational strength through the  dipole moment operator $\bar\psi F\psi $. These are exactly the couplings responsible for the negative contributions in the above table.

For theories containing a massless spin-$3/2$ particle, consistent factorization of the four-particle amplitude forces the presence of the complete supersymmetric spectrum~\cite{McGady:2013sga}. Unlike the gravitino, which comes hand-in-hand with SUSY and which leads to vanishing beta function beyond $\mathcal{N}=1$, the above dimension-five operators can be independently introduced. This motivates us to study the fate of the beta function under these nonminimal couplings in a more general setup.

\subsection[Negative running from nonminimal couplings in non-SUSY theories]{Negative running from nonminimal couplings in non-SUSY \linebreak theories}  \label{sec:running}
Let us now generalize to nonminimally coupled matter fields. We will organize our analysis around specific black hole solutions. In particular, since the dimension-five operators modify the equations of motions, we will only consider cases where the original background is still a solution to the new equations of motion. 

We begin by augmenting Einstein-Maxwell theory with an arbitrary number of scalars and vectors, coupled through $\phi F F$ operators as follows:
\eqa
\mathcal{L} = \mathcal{L}_{\rm EM} - \frac{1}{2}\sum_{i}(\partial\phi_i)^2 + \sum_{i,j,k}g_{ijk}\phi_i F_jF_k.
\eqae
The indices $i$ label the different species. Let us consider a RN black hole charged under one of the photons, say $i=1$; i.e., the only field strength with nontrivial profile is $F_1$, with $F_{1}^2=2(P^2 - Q^2)/r^4$. If $g_{i11}$ is nonvanishing, the scalar equations of motion will be modified by terms proportional to $F_1^2$. Since the RN black hole has a trivial scalar profile, we need to set $P=Q$ and thus consider a dyonic black hole.

Note that with the additional scalars, the dimension-eight operators appearing at one loop now include $(\partial\phi)^2F^2$ and $(\partial\phi)^4$. However, again since the scalar vanishes on the RN solution, such terms will not influence the leading correction to the extremal charge-to-mass ratio. As $F_1^2 =0$, this ratio is only corrected by $a_2(F_1\widetilde{F}_1)^2$. Explicit computation leads to the following modification of the extremal condition:\footnote{Throughout Sec.~\ref{sec:running}, we will suppress the explicit logarithmic factor in the Wilson coefficients, and instead for brevity simply write $a$, $a_1$, or $a_2$ for the coefficient multiplying $\log (s/\mu^2)$, as in \Eq{eq:logs}.}

\eq\label{srnc}
\left.\frac{Q}{M}\right|_{\rm ext}=1+\frac{8}{5Q^2}a_2\, ,
\eqe
where
\eq\label{eq:a2RNphi}
\begin{aligned}
a_2=&\,\frac{1}{12}\sum _{j=2}^{n_p}\left[\left(\sum _{i=1}^{n_s} g _{i1j}^2 - \frac{1}{2}\right)^2  +8  \left(\sum _{i=1}^{n_s} g_{i11} g_{i1j}\right)^2\right] +\frac{4}{3} \left(\sum _{i=1}^{n_s} g_{i11}^2 - \frac{3}{8}\right)^2 \\& + \frac{1}{12}\sum _{\substack{k,l=2\\ k\neq l}}^{n_p} \left(\sum _{i=1}^{n_s} g_{i1k} g_{i1l}\right)^2 + \frac{1}{480}(182+2n_p+n_s),
\end{aligned}
\eqe
writing $n_p$ and $n_s$ for the number of vectors and scalars, respectively.
We see that $a_2$ is given by a sum of positive definite terms.

For vanishing $g_{i11}$, the pure electric RN solution becomes a viable background. Furthermore, since in that case $F\widetilde{F}=-4PQ/r^4 = 0$, we can introduce axion couplings as well. To this end, let us consider:
\be 
\begin{aligned}
\mathcal{L} = \mathcal{L}_{\rm EM} - \frac{1}{2}\sum_{i=2}^{N}\left[\frac{1}{2}F_i^2 + (\partial\phi_i)^2 + (\partial\chi_i)^2\right] + \sum_{i=2}^{N}g_{i1i}\left(\phi_i F_1F_i-\chi_i F_1\widetilde{F}_i\right).\label{eq:counter1}
\end{aligned}
\ee 
For simplicity of the notation, we label the scalar and axion $\phi_i, \chi_i$ with $i=2,\ldots, N$, i.e., we have $N -1$ scalar and axion fields and $N$ vector fields. In $\mathcal{N}=2$ supergravity theories, the scalar and axion combine into the complex scalar in the Maxwell multiplet. Direct computation yields: 
\eq
a_1=a_2=\frac{137}{240}+\sum_{i=2}^N\left(\frac{7}{240}-\frac{g_{i1i}^2}{6}+\frac{g_{i1i}^4}{6}\right).\label{eq:axiondilaton}
\eqe
Note that the minimum occurs when $g_{i1i}^2=1/2$, in which case the expression in parentheses is in fact negative, reaching $-1/80$. 
This special combination of scalar and axion, as well as the value for $g_{i1i}$, leads precisely to the cancellation of the four-graviphoton divergence observed for the Maxwell multiplets in ${\cal N}=2$ supergravity. 
Any deviation would lead to a net positive contribution, which is the case for external matter photons. 
At this special value of $g_{i1i}$, we have $a_1 = a_2 = (140-3N)/240$, so that if we set $N>46$, the beta function would cause the Wilson coefficient to run negative for sufficiently large black holes.

Analogous results also occur for the other dimension-five operator. The electric RN solution also allows us to add the fermion couplings $\bar{\psi}\gamma^{\mu}\gamma^{\nu} F_{\mu\nu}\psi$: 
\eqa
\mathcal{L}=\mathcal{L}_{\rm EM}- \bar{\psi}\slashed{\nabla}\psi + g \bar{\psi} \gamma^{\mu}\gamma^{\nu}F_{\mu\nu}\psi.\label{eq:dim5op}
\eqae
With $N$ copies of such a fermion, we find:
\be
a_1 = a_2 = \frac{137}{240} +  N\left(\frac{1}{160} -\frac{1}{6}g^2 + \frac{2}{3}g^4 \right).\label{eq:a1a2N}
\ee
In $(g,N)$ parameter space, there is a region where the $a_1$ and $a_2$ divergences~\eqref{eq:a1a2N} go negative, as first pointed out in~Ref.~\cite{Charles}. For fixed $N$, \Eq{eq:a1a2N} is minimized for $g^2=1/8$, which is precisely the value for the photino couplings to the graviphoton in extended supergravity theories. At this value we find $a_1 = a_2 = (137-N)/240$, so that the beta function flips sign for $N>137$.
As in the previous case considered above, marginalizing over all $(g,N)$, one finds that this sign flip only occurs in a window of $g \sim {\cal O}(1)$ in Planck units, with the minimal critical $N$ realized at the supergravity value of the coupling.

\subsection{Causality}\label{sec:causality}

The ``wrong-sign'' running of the $F^4$ operator generated by the theories in \Sec{sec:running} would a priori seem to present a problem for causality:
If the $F^4$ operator has negative Wilson coefficient at some (sufficiently-low) energy scale, what is to prevent us from sending signals superluminally using the constructions of Ref.~\cite{Adams:2006sv}?

Let us examine such a thought experiment more closely, in the context of the dipole example in \Eq{eq:dim5op}.
The EFT at scale $E$ will contain a negative $F^4$ correction,
\be 
+\frac{\log(E/m_{*})}{m_{\rm Pl}^4} F^4,\label{eq:F4opneg}
\ee
where $m_{*} \gg E$ is the scale of the UV completion of the dimension-five dipole operator (and so $\log(E/m_{*})<0$ is large).
Suppose we try to exploit this operator to detect superluminality. 
We arrange for an electromagnetic field of strength $B$ in a bubble of size $L$. 
The time advance associated with the $F^{4}$ operator in \Eq{eq:F4opneg} will be
\be
t_{\rm adv} \sim \frac{B^{2}L}{m_{{\rm Pl}}^{4}}\log(m_{*}/E). 
\ee
Meanwhile, the total mass of the bubble is $B^{2}L^{3}$, and the gravitational time delay the signal incurs in crossing the bubble is
\be 
t_{{\rm del}}\sim\frac{B^{2}L^{3}}{m_{{\rm Pl}}^{2}}.
\ee
Thus, in order to obtain net superluminality, we would need $t_{{\rm del}}<t_{{\rm adv}}$, which implies an exponentially small value for the scale $E$:
\be 
\frac{E}{m_{*}}<e^{-m_{{\rm Pl}}^{2}L^{2}}.
\ee
Now, the relevant scale of the EFT is the wavelength of the perturbation of the photon field we are using to send signals within the bubble. Since this must be smaller than the bubble itself, we must have $L>1/E$, so we require
\be 
\frac{E}{m_{{\rm Pl}}}e^{m_{{\rm Pl}}^{2}/E^{2}}<\frac{m_{*}}{m_{{\rm Pl}}}.\label{eq:finaltune}
\ee 
In order for us to treat this theory within QFT, we must have $m_{*}<m_{{\rm Pl}}$. But the function $x\,e^{1/x^{2}}$ is never less than $1$ for positive $x$, so the condition $m_{*}/m_{{\rm Pl}}<1$ is impossible to meet given Eq.~\eqref{eq:finaltune}. 
Thus, we have found that it is not possible for the superluminal time advance to triumph over the gravitational time delay, while allowing the experiment to remain within a consistent EFT.
The same analysis would also apply in the Maxwell multiplet-inspired case considered in \Eq{eq:counter1}.
We note that the fact that the ``wrong-sign'' beta function is set by the Planck scale, leading to the $m_{\rm Pl}$ suppression in \Eq{eq:F4opneg}, was crucial in order for the gravitational time delay to cancel the would-be time advance.

\subsection{UV completion and tuning}\label{sec:tuning}

In addition to the logarithmic running from the beta functions, the Wilson coefficients of the Einstein-Maxwell EFT contain extra, finite contributions that we have so far neglected.
That is, a given $F^4$ coefficient $a$ evaluated at scale $E$ can be expanded as $a_{\rm UV} - a_\beta \log (E/m_{*})$, so that the log-dependent Wilson coefficients we have been computing correspond to $a_\beta$.
The proper scale for $E$ is the Compton wavelength of the horizon $\sim 1/r_{\rm H}$, so the shift in charge-to-mass ratio is, schematically,
\be
\Delta(Q/M) = \frac{1}{r_{\rm H}^{2}}\left[a_{\rm UV}+a_\beta\log(r_{\rm H} m_{*})\right],\label{eq:extremalshifta}
\ee
as depicted in \Fig{fig:qmcurvecounter}. Now, the WGC demands the existence of some state for which $\Delta (Q/M) > 0$, enabling RN black holes to decay.
All extremal black holes with $M>M_0$ can decay provided there exists some $M_{1}<M_{0}$ such that $\Delta (Q/M)|_{M_{1}}>0$. 
For $a_\beta < 0$ as in the theories in \Sec{sec:running}, if $a_{\rm UV} > 0$ there still exists a window of black hole masses in the range $1 \ll r_{\rm H} m_{*} < \exp(-a_{\rm UV}/a_\beta)$ that will have $\Delta(Q/M) > 0$, satisfying the WGC. 
As long as $a_{\rm UV}$ is positive, which we will demonstrate by unitarity in \Sec{sec:squares}, we are guaranteed to have extremal black holes near the threshold region with charge-to-mass ratio greater than 1.

Let us consider more closely the question of the UV completion of the higher-dimension operators in the theories of \Sec{sec:running}.
If we assume a perturbative field-theoretic UV completion, to generate the dimension-five operators of \Sec{sec:running} one must have charged states in the UV. For example, suppose we wish to UV-complete the dipole operator in \Eq{eq:dim5op}. We can imagine a massive complex scalar $\sigma$ with charge $e$ along with a massive fermion $\xi$, both with mass $m_{*}$, that interact with the massless fermions $\psi$ through the Yukawa coupling $y\psi\xi\sigma+{\rm h.c}$.
Then the dipole operator in \Eq{eq:dim5op} is generated at one loop by integrating out $\sigma$ and $\xi$, with Wilson coefficient $g \sim y^2 e/m_{*}$.
To match the example in \Sec{sec:running}, we must have $g\sim 1/m_{\rm Pl}$.
This leaves two possibilities.
Either $y\lesssim 1$ and $e/m_{*} > 1$ in Planck units, in which case $\sigma$ is a particle that already satisfies the WGC, or else $e/m_{*} < 1$, requiring $y \gtrsim 1$, violating perturbativity.
Thus, the existence of a perturbative UV completion of the dipole operator, at least in this example, seems to require the existence of a state in the completion itself that satisfies the WGC.

In the case of negative running, it is intriguing to ask if this implies that in the asymptotic IR one can actually find scales at which the extremal charge-to-mass ratio is less than unity.  
In order to find such an object, one would be required to consider black holes with exponentially large masses.
For example, from a state at $m_{*}$ we might expect threshold corrections in $a_{\rm UV}$ in \Eq{eq:extremalshifta} going like $m_{*}^{-2}$, and as we will discuss in \Sec{sec:squares}, these contributions will be positive.
In the models discussed in \Sec{sec:running}, such terms compete against the negative correction to the charge-to-mass ratio going like $a_\beta \sim -m_{\rm Pl}^{-2}$, which is enhanced by $\log(r_{\rm H}m_{*})$.
For the net extremality correction to be negative, we must consider black holes with horizon size $r_{{\rm H}}>m_{*}^{-1} \exp(m_{{\rm Pl}}^{2}/m_{*}^{2})$.
However, the presence of an exponentially large distance scale suggests sensitivity to the cosmological constant (CC), which we should expect to be nonzero in a SUSY-breaking theory~\cite{Banks:2000fe}.
Let us write the energy density of the CC in a model-independent manner as $m_{\rm Pl}^2 H^2$ for Hubble parameter $H$.
Requiring that the black hole be smaller than the Hubble radius, we must have $r_{\rm H} H < 1$.
Then in order to find an extremal black hole with $Q/M<1$, we must have an extraordinarily exponentially tuned CC, with $H <  m_{*} \exp(-m_{{\rm Pl}}^{2}/m_{*}^{2})$.
To get a sense of how extreme this tuning is, let us input physically motivated numbers.
Minimizing the tuning by taking $m_{*}$ as large as is reasonable, at the GUT scale of $\sim 10^{-2} m_{\rm Pl}$, we would find the requirement that $H \lesssim 10^{-4345}m_{\rm Pl}$, over four thousand orders of magnitude smaller than our own highly-tuned CC of $10^{-120} m_{\rm Pl}^2$.
Even adding back in the scaling with $N\gg 46$ scalars or $N\gg 137$ fermions in $a_\beta$ does not alleviate this tuning. While the enhancement turns the condition on the CC into $H < m_{*} \exp(-m_{\rm Pl}^{2}/Nm_{*}^{2})$, the large number of species renormalizes the effective Planck mass by $\delta m_{\rm Pl}^2 \sim N m_{\rm cutoff}^2$~\cite{Dvali:2007hz,Arkani-Hamed:2005zuc,Dimopoulos:2005ac,Cheung:2014vva}, canceling the $N$-dependence.
Hence, in order to engineer a black hole in tension with the spirit of the WGC, one must also posit either unreasonable tuning of the CC or a desert scenario with no additional degrees of freedom coupling to the photon below the Planck scale.
Such unwelcome features suggest that these models are consigned to the swampland.

\section{Extremality and the action}\label{sec:action}

So far, we have computed the leading-order correction to the charge-to-mass ratio for extremal black holes in the regime of asymptotically large black holes, depicted in yellow in Figs.~\ref{fig:qmcurve} and \ref{fig:qmcurvecounter}. 
Here, the Wilson coefficients were dominated by massless loops, and the sign of the extremality correction was fixed by beta functions.
Motivated by the importance of the contributions from the UV completion in \Sec{sec:tuning}, we now turn toward the threshold regime in green, where the dominant contribution to the Wilson coefficients will be from massive states below the Planck scale.

Before examining particular theories, we first review a profound relationship between the on-shell action and the extremality shift induced by higher-dimension operators shown in Ref.~\cite{dyonic}.
Remarkably, the value of the shift in $Q/M$ at leading order in the EFT is given simply by the value of $\Delta {\cal L}$ itself, evaluated on the on-shell extremal black hole solution in the two-derivative theory.
This fact extends beyond the RN case, generalizing to spinning, multicharge, dyonic, or even dilatonic black holes, and applies for any leading operators $\Delta {\cal L}$, for any number of derivatives.

Consider a Kerr-Newman (KN) black hole with ADM mass $8\pi m/\kappa^2$, electric charge $4\sqrt{2}\pi q/\kappa$, and angular momentum $J=8\pi m a/\kappa^2$, where we reintroduce explicit Planck masses for clarity.
It will be convenient to define the extremality parameter $\zeta = \sqrt{q^2 + a^2}/m$, so that we have $\zeta \in [0,1]$ for physical black holes in Einstein-Maxwell theory. 
The event horizon is located at radius $r_{\rm H} = m(1+\sqrt{1-\zeta^2})$ in Boyer-Lindquist coordinates, so that the extremal case corresponds to $r_{\rm H} = m$, and we define a spin parameter $\nu = a/r_{\rm H}$.
In terms of these parameters, the extremality shift satisfies a beautiful relation,
\be 
\Delta \zeta = \frac{\kappa^2(1+\nu^2)}{8\pi m} \lim_{\zeta \rightarrow 1}\left(\int{\rm d}^3 x \sqrt{-g}\,\Delta{\cal L}|_{\rm KN}\right),\label{eq:Dzeta}
\ee
with the integral evaluated on the KN solution outside the event horizon at fixed $t$.

While we leave the full proof of \Eq{eq:Dzeta} to Refs.~\cite{Cheung:2018cwt,dyonic}, the essential elements of the derivation proceed as follows.
By definition of the horizon, in the extremal limit we mechanically have $\Delta \zeta\propto \Delta g^{rr}$ and $\Delta g^{rr} \propto \Delta r$, the shift in horizon radius at fixed ADM charges. This follows from imposing the condition $g^{rr} = 0$ to locate the horizon, and fixing either the charge or the radial derivative of $g^{rr}$, which vanishes in the extremal limit; see \App{Proof} for details.
In turn, the shift $\Delta S$ in the black hole's Wald entropy from $\Delta {\cal L}$ at fixed charges can be shown to be dominated in the extremal limit by the area shift, so $\Delta r \propto \Delta S$.
Finally, by standard thermodynamic identities in Euclidean quantum gravity, one has a Smarr relation $\Delta S \propto \Delta {\cal L}$~\cite{Reall:2019sah}, which ultimately leads to \Eq{eq:Dzeta}.
The relationship between entropy and extremality was generalized beyond the context of black holes in Ref.~\cite{Goon:2019faz}.

Despite the fact that the relation~\eqref{eq:Dzeta} was derived using a combination of steps in both general relativity and thermodynamics, the final result depends only on the well-defined observables of the on-shell action and extremal charge.
This suggests that an entirely geometrical argument may be possible to mechanically arrive at the extremality/action relation, without appealing to thermodynamics; we leave this possibility to future work.
A particularly beneficial aspect of \Eq{eq:Dzeta} is that it allows the computation of the extremality shift without requiring solving the higher-derivative deformed Einstein equations for the perturbed black hole metric as in Refs.~\cite{Kats:2006xp,Cheung:2018cwt}. 
The agreement of \Eq{eq:Dzeta} with the result obtained via the brute-force method has been explicitly checked for arbitrary four-derivative operators for dyonic RN black holes~\cite{Cheung:2018cwt,dyonic}.

As an example application of \Eq{eq:Dzeta}, consider a theory in which the higher-dimension operators are quartic in the Riemann tensor (cf. type II string theory~\cite{Gross:1986iv}),
\be 
\Delta {\cal L} = \frac{c}{\kappa^2 m_*^6}(R_{\mu\nu\rho\sigma}R^{\mu\nu\rho\sigma})^2  + \frac{\widetilde c}{\kappa^2 m_*^6} (R_{\mu\nu\rho\sigma}\widetilde R^{\mu\nu\rho\sigma})^2,
\ee
where $\widetilde R_{\mu\nu\rho\sigma} = \epsilon_{\mu\nu\alpha\beta}R^{\alpha\beta}_{\;\;\;\;\rho\sigma}/2$. By unitarity~\cite{Bellazzini:2015cra} and causality~\cite{Gruzinov:2006ie}, $c$ and $\widetilde c$ are nonnegative.
For nonspinning black holes, the corrected metric satisfies
\be 
\begin{aligned}
-g_{tt} &= \bar g^{rr} -  \frac{64m^3 c}{715\Lambda^6 r^{14}}\times [2860(11m-8r)r^4  +572 mr^3 z^2(-208m+141r) \\&\hspace{3.8cm}  + 104m^2 r^2 z^4(1593m-925r)  +520m^3 r z^6(-163m+77r) \\&\hspace{3.8cm}  + 12705 m^5 z^8] \\
g^{rr} &=\bar g^{rr} - \frac{64m^3 c}{715\Lambda^6 r^{14}}\times [2860(67m-36r)r^4  + 1716 mr^3 z^2 (-521m+250r) 
\\&\hspace{3.8cm}+ 260m^2 r^2 z^4 (5733 m - 2266r) +5720 m^3 r z^6(-185m+49r)  \\&\hspace{3.8cm}  + 252945 m^5 z^8],
\end{aligned}\label{eq:metricR4}
\ee
where $\bar g^{rr} = 1-(2m/r)+(q^2/r^2)$. The shift in the extremality parameter is
\be
\Delta(q/m) = +\frac{2208 c}{715m_*^6 r_{\rm H}^6}.\label{eq:R4Dz}
\ee
This can be computed either directly from the metric~\eqref{eq:metricR4} or using \Eq{eq:Dzeta}, and we find agreement.

\begin{figure}[t]
\begin{center}
\hspace{-11mm}\includegraphics[width=0.75\columnwidth]{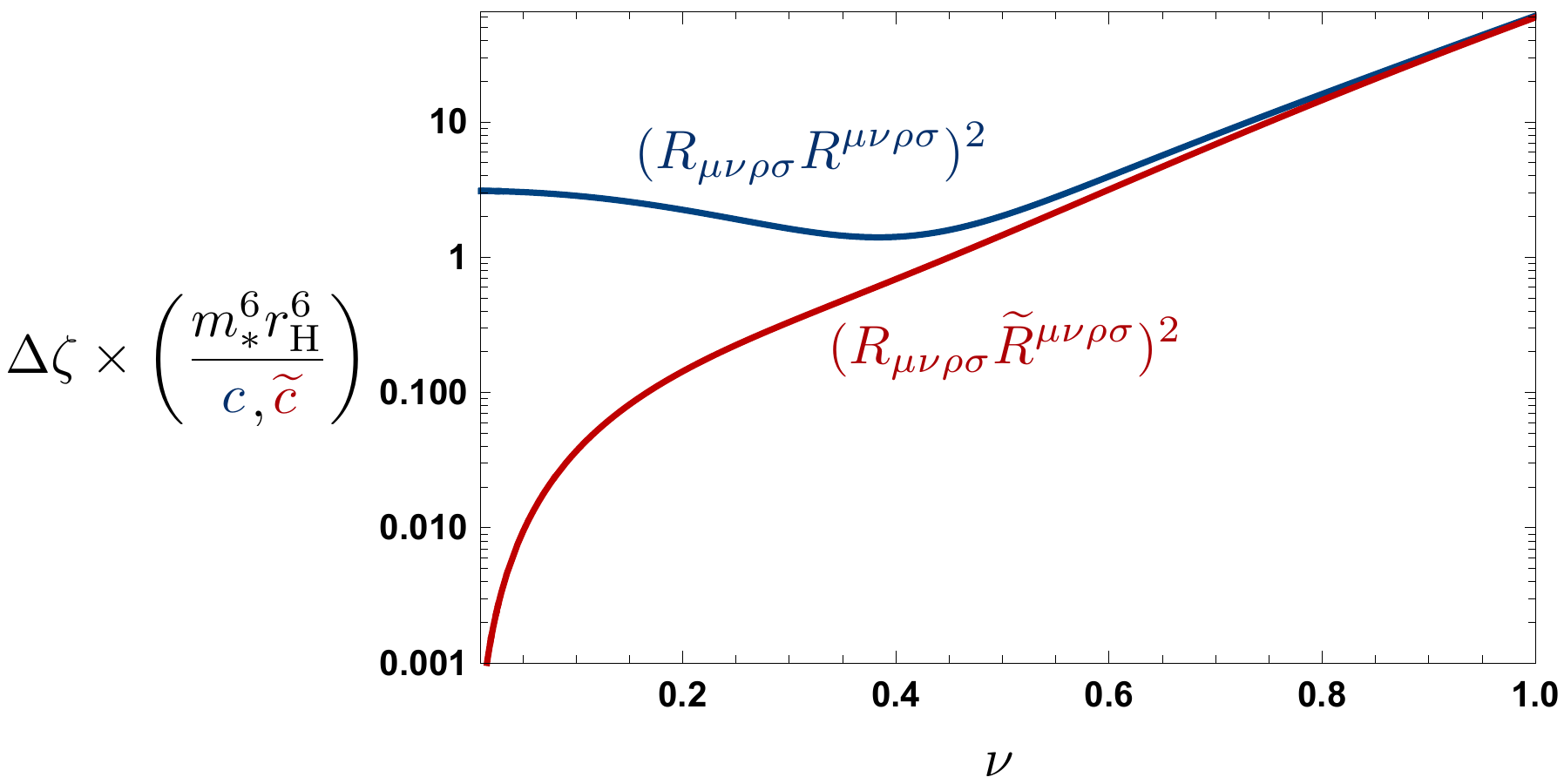}
\end{center}\vspace{-7mm}
\caption{Shift in KN extremality parameter $\zeta$ induced by the quartic Riemann terms.
}
\label{fig:R4}
\end{figure}

Generalizing to spinning charged black holes,  the correction to the KN extremality parameter is
\be 
\begin{aligned}
\Delta\zeta & =\;\;\frac{c}{201600 m_*^{6}r_{\rm H}^{6}\nu^{13}(1+\nu^{2})^{4}}\times\Bigl[3239775\nu^{29}+13046250\nu^{27}+21354690\nu^{25}\\
 & \hspace{2cm} +21664770\nu^{23} +19661192\nu^{21} +14479886\nu^{19}+13943647\nu^{17}\\
 & \hspace{2cm}
 +5093180\nu^{15}+8429455\nu^{13}+20545166\nu^{11}+24409088\nu^{9} \\& \hspace{2cm}
 +11319666\nu^{7}+7270410\nu^{5}+8419530\nu^{3}+3239775\nu\\
 & \hspace{2cm}+315(1+\nu^{2})^{5}\,{\rm arctan}\,\nu\times(10285\nu^{20}-6580\nu^{18}+10734\nu^{16}-2548\nu^{14}\\
 & \hspace{3cm}+709\nu^{12}-10285\nu^{8}+21268\nu^{6}-34566\nu^{4}+21268\nu^{2}-10285)\Bigr]\\
 & \;\;+\frac{\widetilde c}{22400 m_*^{6}r_{\rm H}^{6}\nu^{13}(1+\nu^{2})^{4}}\times\Bigl[225225\nu^{29}+1183350\nu^{27}+2582790\nu^{25} \\&\hspace{2cm} 
 +3052830\nu^{23} +2193344\nu^{21}+1104562\nu^{19}+518289\nu^{17} 
  \\&\hspace{2cm}+1336220\nu^{15}+1603745\nu^{13}+1589258\nu^{11}+1577984\nu^{9}
   \\&\hspace{2cm}+1619958\nu^{7} +1505910\nu^{5}+898590\nu^{3}+225225\nu\\
 & \hspace{2cm}+315(1+\nu^{2})^{5}\,{\rm arctan}\,\nu\times(715\nu^{20}+420\nu^{18}+138\nu^{16}+84\nu^{14}\\
 & \hspace{3cm}+43\nu^{12}-715\nu^{8}+484\nu^{6}-938\nu^{4}+484\nu^{2}-715)\Bigr].
\end{aligned}\label{eq:bigdeltazeta}
\ee
Since $c,\widetilde c>0$, it follows from \Eq{eq:Dzeta} that for all corrected KN black holes (i.e., for all $\nu\in[0,1]$), we have $\Delta \zeta >0$; see \Fig{fig:R4}. As a consistency check of \Eq{eq:Dzeta}, we find that in the $\nu \rightarrow 0$ limit, \Eq{eq:bigdeltazeta} reduces to \Eq{eq:R4Dz} as required.
While $\Delta \zeta > 0$ was not required for the WGC for $\nu \neq 0$, since spinning black holes can shed their angular momentum by emitting Hawking radiation in nonzero orbital angular momentum states, \Eq{eq:bigdeltazeta} shows that such black holes can also decay directly to other spinning black holes with zero orbital angular momentum; such black hole self-sufficiency was also shown for ${\cal O}(\partial^4)$ operators in Ref.~\cite{dyonic}.

\section{Actions and perfect squares}  \label{sec:squares}

In the previous section, we have seen the intriguing relation between the shift of the extremal parameter $\zeta$ and the leading correction for the on-shell action $\Delta {\cal L}$. This prompts us to ask: Is $\Delta {\cal L}$ always positive? Note that from the start, the question is ill-posed due to the simple fact that the leading kinetic term does not have a definite sign.
For concreteness, let us consider a theory with multiple scalars to illustrate this point.  Consider a multiplet of real, shift-symmetric, massless scalars $\phi_{i}, i=1,\ldots,N$, where the action takes the form, 
\eq
\mathcal{L}=-\frac{1}{2}\partial_\mu \phi_i\partial^\mu \phi_i + c_{ijkl} (\partial_\mu \phi_i \partial^\mu \phi_j)(\partial_\nu \phi_k \partial^\nu \phi_l).\label{eq:DLmultiphi}
\eqe   
A particular solution to the two derivative part of the action can take the form
\eq
\partial_\mu \phi_i=v_i f_\mu,\label{eq:vf}
\eqe
where $v_i$ is some ``flavor'' vector and $f$ is a spacetime-dependent four-vector. Solutions obeying this ansatz transform simply under the ${\rm O}(N)$ symmetry of the two-derivative action. On such a solution, the kinetic term is $f^2 v^2$ and, depending on the signature of the vector $f$, can take any sign.

Similar behavior can be found for the leading four-derivative correction. The coupling $c_{ijkl}$ by definition satisfies  
\be 
c_{ijkl}=c_{klij}=c_{jikl}=c_{ijlk}.\label{eq:sym}
\ee
This is reducible and we can further decompose it by symmetrizing and antisymmetrizing on $j,k$, while continuing to respect the symmetry in \Eq{eq:sym}. Doing so, we write $c_{ijkl} = c^S_{ijkl}{+}c^A_{ijkl}$, defining
\be 
\begin{aligned}
c^S_{ijkl} &= c_{(ijkl)} \\
&=\frac{1}{3}(c_{ijkl}+c_{ikjl}+c_{ilkj})\\
c^A_{ijkl} &= c_{i[jk]l}+\text{symmetrize on Eq. }\eqref{eq:sym}\\
&= \frac{1}{3}(2c_{ijkl}-c_{ikjl}-c_{ilkj}),
\end{aligned}\label{eq:cSA}
\ee
where we use round and square brackets to denote normalized (anti-)symmetrization. 
We can then decompose $\Delta {\cal L}$ as
\be
\begin{aligned}
\Delta{\cal L} &= \Delta{\cal L}^S + \Delta{\cal L}^A \\
\Delta {\cal L}^S &= c_{ijkl}^S (\partial_\mu \phi_i \partial^\mu \phi_j)(\partial_\nu \phi_k \partial^\nu \phi_l)\\
\Delta {\cal L}^A &= c_{ijkl}^A (\partial_\mu \phi_i \partial^\mu \phi_j)(\partial_\nu \phi_k \partial^\nu \phi_l).\label{eq:DLphidecomp} 
\end{aligned}
\ee
Now since $\Delta {\cal L}^A=0$ on \Eq{eq:vf}, this means deformation around this background can take any sign irrespective of the coefficient  $c^{A}_{ijkl}$, and thus it not possible for $\Delta {\cal L}^A$ to have a definite sign in general.
On the other hand, $\Delta {\cal L}^S$ does not vanish on \Eq{eq:vf} and therefore has a chance at being positive on such backgrounds.
Remarkably, precisely these terms, that have the possibility of being positive on factorized backgrounds of the form \eqref{eq:vf}, are in fact guaranteed to be so from unitarity and causality. (However, neither $\Delta {\cal L}^S$ nor $\Delta {\cal L}^A$ is positive for complete arbitrary field configurations.) As we will see, this conclusion comes from the dispersive representation for $c_{ijkl}$.

These statements can be extended to other operators quartic in field strengths, e.g., those involving $N$ ${\rm U}(1)$ gauge fields. Importantly, black hole backgrounds are those for which the analogue of $\Delta {\cal L}^A=0$ and the correction to the extremal parameter is solely determined from  $\Delta {\cal L}^S$, which we will find is positive for such backgrounds. By virtue of the action/extremality relation in \Eq{eq:Dzeta}, this implies that the shift in charge-to-mass ratio is positive in the green threshold region of Figs.~\ref{fig:qmcurve} and \ref{fig:qmcurvecounter}.
Hence, the finite corrections from the UV states themselves imply that black holes satisfy the WGC.
Throughout this section, we will restrict consideration to actions starting at quartic order in the massless modes.\footnote{This is consistent from an EFT perspective and corresponds to a UV completion that does not modify the three-point amplitudes of the low-energy modes at leading order.
In particular, the $R_{\mu\nu\rho\sigma}F^{\mu\nu}F^{\rho\sigma}$ operator will be excluded.}

\subsection{Causality and unitarity}\label{sec:generalizedunitarity}
Taking the Wilson coefficients in \Eq{eq:DLmultiphi} to be generated by integrating out massive particles parametrically lighter than the Planck scale, we can constrain $c_{ijkl}$ using dispersion relations.
The most general set of bounds on $c_{ijkl}$ comes from making use of unitarity in the form of the generalized optical theorem~\cite{ZZ}. In particular, such bounds can be stricter than any dispersive relation one obtains from elastic scattering of superpositions of states.

Define $2\,M^{ijkl}\,{=}\,{\rm d}^2 M_{ij\rightarrow kl}(s,t{=}0)/{\rm d}s^2$, where we write $M_{ij\rightarrow kl}(s,t)$ for the amplitude for $\phi_i \phi_j \rightarrow \phi_k \phi_l$, working in cyclic formalism so that elastic scattering would correspond to $i=l$ and $j=k$. By the standard construction of the analytic dispersion relation~\cite{Adams:2006sv},
\be 
\hspace{-2mm}M^{ijkl}= \frac{1}{2\pi i} \int_{0}^{\infty}\frac{{\rm d}s}{s^{3}}\left[{\rm disc}\,M_{ij\rightarrow kl}(s)+{\rm disc}\,M_{ik\rightarrow jl}(s)\right].
\ee
Though we write the lower limit of integration as zero, in practice we could instead start the integral---and evaluate the Wilson coefficients---at a finite threshold scale below the mass of the UV states generating $\Delta {\cal L}$. 
Now, define the three-point amplitude for $\phi_{i}(p_1)\phi_{j}(p_2)\rightarrow X$, where $X$ is any final state, and write out its real and imaginary parts as ${\cal A}_{ij\rightarrow X}=m_{R_{X}}^{ij}+i\,m_{I_{X}}^{ij}$.
We thus have
\be 
\begin{aligned}
M^{ijkl}&=\frac{1}{\pi}\int_{0}^{\infty}\frac{{\rm d}s}{s^{3}}\sum_{X}\Big(m_{R_{X}}^{ij}m_{R_{X}}^{kl} +m_{I_{X}}^{ij}m_{I_{X}}^{kl} +m_{R_{X}}^{ik}m_{R_{X}}^{lj} +m_{I_{X}}^{ik}m_{I_{X}}^{lj}\Big),
\end{aligned}\label{eq:generaldisp}
\ee
where we drop the boundary term at infinity by requiring sufficiently well-behaved scaling ($\lesssim s^2$) of the forward amplitude at large momentum~\cite{Adams:2006sv}.
Now, $m_{R_{X}}^{ij}$ and $m_{I_{X}}^{ij}$ are some unknown, arbitrary, real-valued matrices. 
Thus, the $M^{ijkl}$ allowed by unitarity are given by the full set of positive sums of $m^{ij}m^{kl}+m^{ik}m^{lj}$, where the $m^{ij}$ are real $N$-by-$N$ matrices. 
The set of quadratic outer products of matrices defines a cone ${\cal C}$: given any two $M^{ijkl}$ in ${\cal C}$, say $M_{1}$ and $M_{2}$, one has $\lambda_{1}M_{1}+\lambda_{2}M_{2}\in{\cal C}$ for all $\lambda_{1},\lambda_{2}>0$.

Computing the amplitudes from \Eq{eq:DLmultiphi}, we find $M_{ijkl}=2(c_{ijkl} + c_{ikjl})$, so the result of the generalized optical theorem is that
\be 
c_{ijkl} + c_{ikjl} = \sum_m (m^{ij}m^{kl}+ m^{ik} m^{lj}).\label{eq:cc}
\ee
Using the symmetries of $c_{ijkl}$ in \Eq{eq:sym} as described in \App{app:cc}, we can show that \Eq{eq:cc} is equivalent to an elegant expression for $c_{ijkl}$ alone,
\be 
c_{ijkl}=\sum_{m}\left(m^{(ij)}m^{(kl)}+m^{[il]}m^{[kj]}+m^{[ik]}m^{[lj]}\right),\label{eq:ccfinal}
\ee
which immediately leads to 
\eq
c^S_{ijkl}=\sum_{m}\,m^{(ij}m^{kl)},
\eqe
where we have symmetrized over all indices. We note that unitarity generates more bounds in \Eq{eq:ccfinal} than obtained by considering elastic scattering of arbitrary superpositions of scalars as in Ref.~\cite{Andriolo:2020lul}. 

For arbitrary $\phi_i$ backgrounds, using the results of Eq.~\eqref{eq:ccfinal} one finds that neither $\Delta {\cal L}^S$ nor $\Delta {\cal L}^A$ is positive in general. 
Defining 
\eq
T^{ij}=\partial_\mu \phi_i \partial^\mu \phi_j,
\eqe 
we can consider the cases in which $T_{ij}T_{kl}$ obeys the symmetries of $c^S$ or $c^A$ as defined in \Eq{eq:cSA},
\be
\begin{aligned}
(T_{ij}T_{kl})^S &= \frac{1}{3}(T_{ij}T_{kl} + T_{ik}T_{jl} + T_{il}T_{kj}) \\
(T_{ij}T_{kl})^A &= \frac{1}{3}(2T_{ij}T_{kl} - T_{ik}T_{jl} - T_{il}T_{kj}).
\end{aligned} 
\ee
First, note that $(TT)^S c^A=(TT)^A c^S=0$. Thus, depending on the symmetry property of $(TT)$, either $\Delta \mathcal{L}^S$ or $\Delta \mathcal{L}^A$ can be zero, which means that it cannot have a definite sign irrespective of the sign of $c^S$ or $c^A$. However if $(TT)$ has the symmetry property that kills $c^S$, i.e., if $(TT) = (TT)^A$, then $(TT)c^A$ reduces to $\sum_m \left(T_{ij}m^{ij}\right)^2 > 0$ and vice versa for $S\leftrightarrow A$. That is, if $(TT)^{S,A} = 0$ then $(TT)c^{A,S}$ is positive. 
In particular, for backgrounds of the form in Eq.~\eqref{eq:vf}, we find:
\be 
\Delta{\cal L}^S=(f^2)^2\sum_{m}(v\cdot m\cdot v)^2>0.\label{eq:DLsquare}
\ee
The ansatz \eqref{eq:vf} obeys a no-hair theorem of sorts, in that we have required $f_\mu$ to be independent of the direction of $v_i$ under which the solution is ``charged'' in flavor space.
Indeed, such a solution would govern the scalar profile of a generalization of a dilatonic black hole to a theory in which the dilaton is replaced by a multiplet $\phi_i$.

The positivity statement of \Eq{eq:DLsquare} can be generalized to any theory where the quartic amplitude at leading order goes like $s^2$, with arbitrary states replacing the scalar multiplet.
For example, consider a CP-conserving EFT with gauge group $\Pi_{i=1}^N {\rm U}(1)_i$:
\be 
\Delta {\cal L} = c_{ijkl}(F^{i}F^{j})(F^{k}F^{l})+\widetilde c_{ijkl}(F^{i}\widetilde{F}^{j})(F^{k}\widetilde{F}^{l}).\label{eq:DLF4multi}
\ee
If we scatter photons of arbitrary flavor with parallel or perpendicular linear polarizations, only one of the two operators (the first or second, respectively) contributes to the forward amplitude. 
Writing the real or imaginary parts of the amplitudes for $\gamma_i(p_1)\gamma_j(p_2)\rightarrow X$ with parallel (perpendicular) linear polarizations as $m^{ij}$ ($\widetilde m^{ij}$), \Eq{eq:ccfinal} immediately generalizes for \Eq{eq:DLF4multi}:
\be
\begin{aligned}
c_{ijkl}&=\sum_{m}\left(m^{(ij)}m^{(kl)}+m^{[il]}m^{[kj]}+m^{[ik]}m^{[lj]}\right)\\
\widetilde c_{ijkl}&=\sum_{\tilde m}\left(\widetilde m^{(ij)}\widetilde m^{(kl)}+\widetilde m^{[il]}\widetilde m^{[kj]}+\widetilde m^{[ik]}\widetilde m^{[lj]}\right).
\end{aligned} \label{eq:ccfinalgauge}
\ee
Consider a gauge field background that factorizes analogously to \Eq{eq:vf},
\be 
F^i = Q_i\, f+ P_i\, \star\! f,\label{eq:Fform}
\ee
where $Q_i$ and $P_i$ are arbitrary $N$-component vectors, $f$ is an arbitrary spacetime-dependent two-form with components $f_{\mu\nu}$,  and $\star f$ is its Hodge dual with components that we write as $\widetilde f_{\mu\nu} = \epsilon_{\mu\nu\rho\sigma} f^{\rho\sigma}/2$.
(Note in particular that the field strength for an arbitrary dyonic black hole charged under this multi-${\rm U}(1)$ theory is a particular example of \Eq{eq:Fform}, with $F^{i}=\frac{Q_{i}}{r^{2}}\,{\rm d}t\wedge{\rm d}r+P_{i}\sin\theta\,{\rm d}\theta\wedge{\rm d}\phi$.)
From \Eq{eq:ccfinalgauge}, unitarity implies that for solutions with the form \eqref{eq:Fform}, $\Delta {\cal L}$ is positive:
\be 
\begin{aligned}
\Delta{\cal L } &=\phantom{+}\sum_{m}\Big\{\left[(Q\cdot m\cdot Q-P\cdot m\cdot P)(f^{2})+(Q\cdot m\cdot P+Q\cdot P\cdot m)(f\widetilde{f})\right]^{2} \\& \qquad\qquad + \left[(Q\cdot m\cdot P)-(P\cdot m\cdot Q)\right]^{2}\left[(f^{2})^{2}+(f\widetilde{f})^{2}\right] \Big\} 
\\&\phantom{=}+\sum_{\tilde m}\Big\{\left[(P\cdot\widetilde{m}\cdot Q+Q\cdot\widetilde{m}\cdot P)(f^{2})-(Q\cdot\widetilde{m}\cdot Q-P\cdot\widetilde{m}\cdot P)(f\widetilde{f})\right]^{2}\\&\qquad\qquad+\left[(Q\cdot\widetilde{m}\cdot P)-(P\cdot\widetilde{m}\cdot Q)\right]^{2}\left[(f^{2})^{2}+(f\widetilde{f})^{2}\right]\Big\} 
\\& > 0,
\end{aligned}\label{eq:gaugepositivity}
\ee
writing $P\cdot m\cdot Q$ for $m^{ij}P_{i}Q_{j}$, $P\cdot\widetilde{m}\cdot Q$ for $\widetilde{m}^{ij}P_{i}Q_{j}$, etc.
(Indeed, writing the analogues of $c^S$ and $c^A$ in \Eq{eq:cSA} for \Eq{eq:ccfinalgauge}, both the symmetric and antisymmetric parts contribute to \Eq{eq:gaugepositivity}, but the antisymmetric part drops out if either $Q_i$ or $P_i$ vanishes in \Eq{eq:Fform}.)
In addition to positivity, we further expect that $\Delta {\cal L}$ is convex as a functional of the fields, as discussed in \App{app:convex}.

The result of \Eq{eq:Dzeta} and the argument for positivity of $\Delta {\cal L}$ on black hole solutions---for quartic and higher contact operators in unitary UV completions---is that for any such deformation of Einstein-Maxwell theory, including beyond the leading four-derivative terms, the extremality parameter receives a positive correction. 
In particular, higher-derivative corrections of the form $F^4$, $F^8$, $R^4$, $R^2 F^2$, etc. should contribute positively to the extremal charge-to-mass ratio, thus allowing black holes themselves to be the states demanded by the WGC.
While the sign of the action can be ambiguous in the case of cubic terms---particularly the $R_{\mu\nu\rho\sigma}F^{\mu\nu}F^{\rho\sigma}$ operator for extremal KN black holes with small spin---we expect that $\Delta \zeta$ will be positive for any stable background~\cite{dyonic}. As noted previously, it is consistent from an EFT perspective to consider a theory in which the first corrections in $\Delta {\cal L}$ occur at quartic order and higher.

\subsection{Scalar examples}
Let us consider some particular example EFTs with tree-level completions in more detail.
Starting with the Einstein-Maxwell-dilaton Lagrangian,
\eqa
\mathcal{L}_{\rm EMD}=R-2(\partial\phi)^2-\frac{1}{2}e^{-2\lambda\phi}F^2,\label{eq:EMDL}
\eqae
the resulting charge-to-mass ratio of the extremal magnetic dilaton black hole (the GHS solution~\cite{GHS} for arbitrary dilaton coupling $\lambda$) is modified by dimension-eight operators,\footnote{See \App{GHSSol} for details. Here we are assuming that the operators are dressed with appropriate powers of exponential dilaton factors  such that in the string frame, the Lagrangian has a universal dilaton factor.} 
\be 
\begin{aligned}
\Delta {\cal L} = a_1e^{-6\lambda\phi}(F^2)^2  + b_1e^{-4\lambda\phi}(\partial\phi)^2 F^2 + c\,e^{-2\lambda\phi}(\partial\phi)^4,
\end{aligned}\label{eq:EMDEFT}
\ee
in terms of which we find:
\be 
\begin{aligned}
&\left.\frac{P}{\sqrt{2(1 + \lambda^2)}M}\right|_{\rm ext} =1 + \frac{16(1 + \lambda^2)^2a_1 + 4\lambda^2(1 + \lambda^2)b_1 + \lambda^4c}{10(1+\lambda^2)^4P^2}.\label{eq:Deltazdilaton}
\end{aligned}
\ee

Let us now give examples for $(a_1,b_1,c)$ by considering a massive scalar $X$ coupled to photons and dilatons $\phi$ as
\eq
f_1XF^2,\quad f_2 X(\partial \phi )^2.
\eqe 
In the low-energy theory that emerges from integrating out the massive scalar at tree level, we find:
\be 
a_1=\frac{f_1^2}{8m_X^2},\;\;b_1=\frac{f_1 f_2}{4m_X^2},\;\; c=\frac{f_2^2}{8m_X^2}. \label{eq:dilatonabc}
\ee
Importantly, note that $b_1$ can be negative. However, we find the charge-to-mass ratio from Eq.~\eqref{eq:Deltazdilaton} is now given by
\be \label{Answ}
\left.\frac{P}{\sqrt{2(1+\lambda^2)}M}\right|_{\rm ext}=1+\frac{[4(1+\lambda^2)f_1+\lambda^2f_2]^2}{80(1+\lambda^2)^4P^2m_X^2}.
\ee
In other words, the sign-ambiguous term $f_1f_2$ gets combined with others to form a perfect square. 
Thus we see that, in this simple scalar example, the individual coefficients of the four-derivative operators may not have a definite sign, but their contribution to the charge-to-mass ratio does. 

The perfect square above is reflecting the fact that the on-shell action itself is a perfect square. Substituting \Eq{eq:dilatonabc} into $\Delta {\cal L}$, we have:
\be 
\Delta {\cal L} = \frac{e^{-2\lambda\phi}}{8m_X^2} \left[ f_1 e^{-2\lambda\phi}F^2 + f_2 (\partial\phi)^2\right]^2.
\ee
Indeed, substituting the GHS solution and integrating from the horizon to infinity, one reproduces \Eq{Answ}. Remarkably, we find that the extremality/action relation in \Eq{eq:Dzeta} also holds for the higher-derivative-corrected GHS black hole.
That is (in appropriate units of mass and Newton's constant) one finds by explicit calculation that the the correction to the extremal charge-to-mass ratio computed directly from the equations of motion in \Eq{eq:Deltazdilaton} and \App{GHSSol} actually satisfies the $\Delta \zeta \sim \Delta {\cal L}$ relation in \Eq{eq:Dzeta} for arbitrary dilaton coupling constant.
This is surprising, since extremal GHS black holes are qualitatively very different from extremal KN---in particular, the former have vanishing entropy and area---and the mechanics of the proof of \Eq{eq:Dzeta} above relied on the perturbation of a nonzero horizon area in order to relate the charge-to-mass shift to the on-shell action.
However, in string frame, the GHS black hole has recently been shown to exhibit an ${\rm AdS}_2\times S^2$ geometry~\cite{Porfyriadis:2021zfb}, and in Einstein frame the higher-derivative terms can themselves similarly regularize the singular extremal GHS limit with a Bertotti-Robinson-like spacetime akin to near-horizon RN with nonzero entropy~\cite{Herdeiro:2021gbw}.
Such results may provide a path to understanding the empirical observation that the extremality/action relation in \Eq{eq:Dzeta} nonetheless holds for a GHS black hole.

Similar perfect-square behavior occurs in a related example, where we allow operators of different mass dimensions to contribute. 
While the signs of the couplings of the $F^4$ and quartic Riemann operators are constrained by analyticity of scattering amplitudes~\cite{Adams:2006sv,Cheung:2014ega,Bellazzini:2015cra}, the signs of six-derivative operators of the form $R^2 F^2$ are not constrained without invoking additional assumptions.
This leads to a natural question: In EFTs with nonzero $F^4$ and $R^4$ terms, derived from a well-defined UV completion, can the couplings be such that the indefinite-sign $R^2 F^2$ operators overwhelm the definite-sign operators of higher and lower dimension, making the sign of the shift in charge-to-mass ratio indefinite? 
As a concrete example, consider Einstein-Maxwell theory with a massive scalar $X$ coupled to $F^2$ and the various $R^2$ terms as:
\be 
{\cal L}_{{\rm UV}}=\frac{R}{2\kappa^{2}}-\frac{1}{4}F^{2}-\frac{1}{2}(\partial X)^{2}-\frac{1}{2}m_{X}^{2}X^{2} + g_X X \left(a_{1}R^{2} {+} a_{2}R_{\mu\nu}R^{\mu\nu} {+} a_{3}R_{\mu\nu\rho\sigma}R^{\mu\nu\rho\sigma} {+} \epsilon bF^{2}\right),
\ee
where $a_{1,2,3}$ and $b$ are ${\cal O}(1)$ constants in Planck units, $\epsilon$ is a unitless parameter that we can tune, and $g_X \ll 1$ is a small unitless coupling. Integrating out $X$, we have an EFT where $\Delta {\cal L}$ is given by 
\be
\frac{g_X^2}{2m_{X}^{2}}\left(a_{1}R^{2} +a_{2}R_{ab}R^{ab} +a_{3}R_{abcd}R^{abcd}  + b \epsilon F^2 \right)^{2}.\label{eq:DLphi}
\ee
When $\epsilon \sim 1$, the $F^4$ terms in \Eq{eq:DLphi} dominate, while when $\epsilon \ll 1$, the $R^4$ terms dominate, but for intermediate $\epsilon$, one might imagine that the $R^2 F^2$ terms, with Wilson coefficients of indefinite sign, will dominate and spoil the positivity of the extremality shift of RN black holes.
However, computing the corrected metric explicitly for an arbitrary dyonic black hole, we find that the shift in the charge-to-mass ratio can be written in a form that manifestly satisfies the WGC:
\be
\begin{aligned}
\left.\frac{\sqrt{Q^2 + P^2}}{\sqrt{2}M}\right|_{\rm ext}  & =1+\frac{\kappa^{4}g_X^2}{m^{6}m_{X}^{2}} \left[\frac{46656}{614185}\left(\frac{95}{108}a_{2}+a_{3}\right)^{2}\right. \\&\qquad\qquad\qquad +\frac{2352}{9449} \left(  a_{2} + \frac{17}{7}a_{3} - \frac{9449}{5292}\frac{b\epsilon m^{2}}{\kappa^{2}}\gamma \right)^{2}\\&\qquad\qquad\qquad +\left.\frac{383}{59535}\frac{b^{2}\epsilon^{2}m^{4}}{\kappa^{4}}\gamma^2\right],
\end{aligned}\label{eq:Dzcompetition}
\ee
writing $\gamma = (Q^2 - P^2)/(Q^2 + P^2)$. 
Here, we must take the coupling $g_X$ small, so that second-order back reaction effects from two insertions of the $F^4$ operator (which would contribute $\propto g_X^4$) do not compete with the $R^2 F^2$ or $R^4$ effects in \Eq{eq:Dzcompetition} ($\propto g_X^2$).
While this conspiracy of operators of different mass dimension to arrange themselves such that the charge-to-mass ratio increases in a well-defined UV completion may seem mysterious at the level of \Eq{eq:Dzcompetition}, it is directly connected to the sum-of-squares form of the higher-derivative terms in the action that arises as a consequence of unitarity.

\section{Outlook}\label{sec:outlook}
In this paper, we have investigated the behavior of corrections to the extremality condition for black holes in two regimes: the asymptotic IR and the threshold regime.
In the former, the charge-to-mass ratio of an extremal black hole is quantum mechanically modified by $T_{\mu\nu}T^{\mu\nu}$ corrections to the EFT induced by loops of massless states.
We show that these corrections are always positive, except in two special cases involving large numbers of massless particles with specific nonminimal, Planck-suppressed couplings; in these cases, we argued that, despite appearances, causality is not in danger and the WGC can be preserved by threshold effects.

In the threshold regime, where finite contributions to the EFT from integrating out massive states below $m_{\rm Pl}$ are dominant, we reviewed a result of Refs.~\cite{Cheung:2018cwt,dyonic}, giving a profound connection between the extremality correction and the on-shell EFT action.
We performed nontrivial checks of this result using both GHS black holes and quartic Riemann operators.
We then employed locality and unitarity---in the form of analytic dispersion relations and the generalized optical theorem---to show that healthy EFTs with higher-derivative terms at quartic order or higher must have an action in the form of a perfect square.
We proved this statement explicitly in the case of four-derivative operators in theories with an arbitrary number of scalars or photons.

Unitarity and the WGC are fundamentally interconnected both by their overlapping consequences for the positive-definite form of the on-shell Lagrangian and their connections to the running of EFT operators. Remaining open questions include more closely studying the fate of the nonminimally coupled, nonsupersymmetric theories with $>137$ fermions or $>46$ bosons, in order to more conclusively decide whether they belong to the landscape or the swampland, and understanding whether unitarity and causality can put practically useful bounds on the leading cubic higher-dimension operators. More broadly, the application of constraints from causality and unitarity to prove or sharpen other swampland conjectures is an interesting direction for future work.
The unexpected utility of the classical Lagrangian as a means of computing the black hole extremality bound in general theories suggests that other applications of the on-shell action could also be worthy of investigation. 

Often, discussions of swampland criteria emphasize their nature as ``intrinsically quantum gravitational'' statements, with no field-theoretic avatars. But at least in the context of the WGC, we have seen that the consequences of ``special theories that provide healthy UV completions of quantum gravity'' and ``universal constraints from unitarity and causality'' are instead part of a continuum of statements. If we look at the $Q/M$ curve as a function of $M$ for theories with any amount of supersymmetry where the asymptotic flat-space region can be parametrically realized, we find at all masses that there are massive states above the extremal line. For masses larger than $M \gtrsim m_{\rm Pl}^2/m_*$, where the black hole solutions can be trusted in the effective field theory, the ``reason'' for this is given by the constraints from causality, unitarity, and dispersion relations. For exponentially large $M \gtrsim (m_{\rm Pl}^2/m_*)\exp(m_{\rm Pl}^p/m_*^p)$, it is the logarithmic running of the operators that guarantees the negative shift. Of course, any statement for masses $M \lesssim m_{\rm Pl}$ depends on detailed knowledge of an actual UV completion, and it is here that the beautiful facts about the existence of {\it light} states satisfying the WGC in all known examples in string theory are relevant. But these three regimes appear to be united in suggesting that $Q/M$ curve is convex as a function of $M$, approaching ``$1$'' monotonically from above. Such monotonicity properties were already noted in the spectrum of charged states in the heterotic string in Ref.~\cite{ArkaniHamed:2006dz}. More recently, an interesting connection between the convexity of the spectrum of charged operators in AdS has been discovered in Ref.~\cite{Ofer}. 

Studies of the WGC have seen that many  fundamental aspects of physics, ranging from unitarity and causality to black hole decay, are intertwined and mutually enforcing.
The ultimate nature and full consequences of these connections in unraveling the workings of quantum gravity, and their ability to make nontrivial predictions for real-world physics, remains as a worthy challenge for future investigation.

\pagebreak

\begin{center} 
{\bf Acknowledgments}
\end{center}
\noindent 
We thank Clifford Cheung, Hirosi Ooguri, Timothy Trott, and Cumrun Vafa for useful discussions and comments. 
{N.A-H.} is supported by DOE grant DE-SC0009988. Y.-t.H. and J.-Y.L. are supported by MoST grant 109-2112-M-002 -020 -MY3. 
G.N.R. is supported at the Kavli Institute for Theoretical Physics by the Simons Foundation (Grant~No.~216179) and the National Science Foundation (Grant~No.~NSF PHY-1748958) and at the University of California, Santa Barbara by the Fundamental Physics Fellowship.

\appendix

\section{Scalar bubble coefficients}\label{app:bubble}
In this appendix, we give a brief review of the extraction of bubble coefficients from unitarity cuts, mainly following the construction in Refs.~\cite{ArkaniHamed:2008gz, Britto:2005ha}. One starts with the $s$-channel cut of the one-loop amplitude, which is given by the product of two tree amplitudes on both sides of the cut:
$$\includegraphics[scale=0.4]{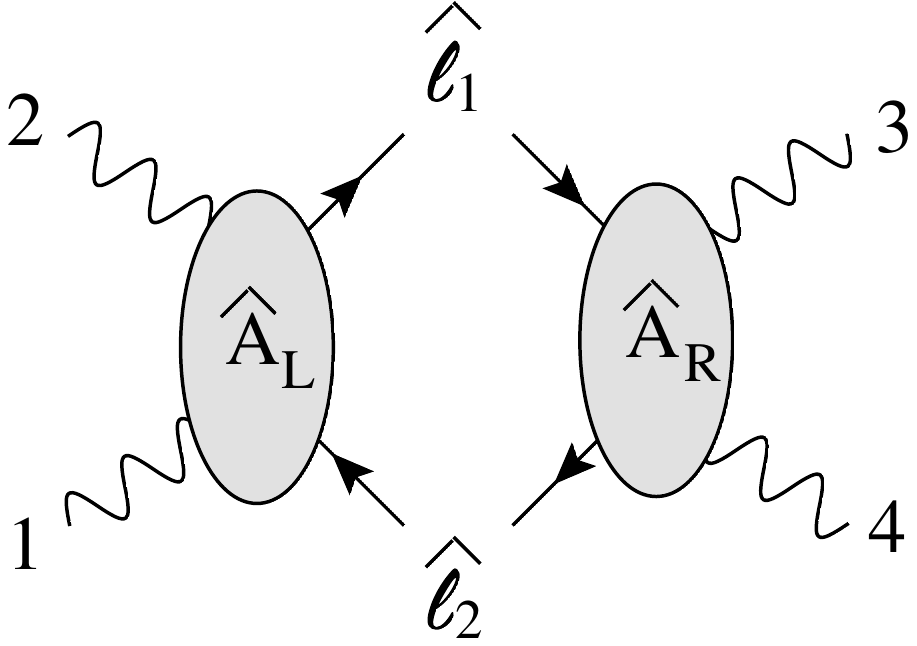}$$
The loop momenta $(\hat{\ell}_1,\hat{\ell}_2)$ are deformed by a complex parameter $z$,
\eq
\hat{\ell}_1=\ell_1+q z,\quad \hat{\ell}_2=\ell_2-q z,
\eqe
in such a way that momentum conservation and null conditions are satisfied.
The corresponding bubble coefficient $c_s$ can then be calculated as
 \begin{eqnarray}\label{bubble1}
  c_s =
  \frac{1}{(2\pi i)^2}
  \int {\rm d} \text{LIPS}[\ell_1,\ell_2]
  \int_\mathcal{C}\frac{{\rm d}z}{z}\sum_\text{state sum}
  \hat{A}_L^\text{tree}\big( \hat{\ell}_1,\hat{\ell}_2\big)\,
  \hat{A}_R^\text{tree}\big( \hat{\ell}_1,\hat{\ell}_2\big) \,,~~~~
\end{eqnarray}
where ${\rm d}{\rm LIPS}={\rm d}^4\ell_1 {\rm d}^4\ell_2 \,\delta^{(+)}(\ell_1^2)\,\delta^{(+)}(\ell_2^2)\,\delta^{4}(-\ell_1+\ell_2+K)$ and $K$ is the momentum going out of the right subamplitude. For the $s$-channel cut here, $K=p_3+p_4$. The product $\hat{A}_L \hat{A}_R$ has poles in $z$ at finite values and at infinity, with the former corresponding to extra loop propagators going on shell, i.e., the contributions from the scalar triangle and box integrals. Thus, the scalar bubble coefficient is given solely by the pole at infinity. After parameterizing $\ell_1=\ell_2+K, \ell_2=t\lambda\tilde\lambda$, integration can be written as
\eq
\int {\rm d} \text{LIPS}=K^2\int\frac{\AB{\lambda {\rm d}\lambda}\SB{\tilde{\lambda} {\rm d}\tilde{\lambda}}}{\ASB{\lambda | K | \tilde{\lambda}}^2}.
\eqe 
The integration of ${\rm d}$LIPS has become contour integration along $\tilde{\lambda}=\bar{\lambda}$.
With momentum conservation, the residue at infinity can be expanded as 
\eq
\begin{split}
\underset{z\to\infty}{\rm Res}\left[\frac{1}{z}
  \hat{A}_L^\text{tree}\big( \hat{\ell}_1,\hat{\ell}_2\big)\,
  \hat{A}_R^\text{tree}\big( \hat{\ell}_1,\hat{\ell}_2\big)\right]=\sum_i a_i \frac{\prod_j\AB{\lambda A_j}}{\prod_k\AB{\lambda B_k}}\prod_l\frac{\SB{\tilde{\lambda} C_l}}{\ASB{\lambda | K | \tilde{\lambda}}}\mathcal{F}(\lambda_e,\tilde{\lambda}_e),
\end{split}
\eqe
where $\mathcal{F}(\lambda_e,\tilde{\lambda}_e)$ is a function of the external spinors. Using the identity
\eq
\frac{1}{{\ASB{\lambda | K | \tilde{\lambda}}}^2}\prod_{l=1}^{n} \frac{\SB{\tilde{\lambda} C_l}}{\ASB{\lambda | K | \tilde{\lambda}}} = \frac{1}{(n+1)!}\prod_{l=1}^{n} \left( \lambda_C^{\dot{a}}\frac{\partial}{\partial \tilde{P}^{\dot{a}}} \right)\frac{1}{\SB{\tilde{P} \tilde{\lambda}}^2},
\eqe
where $\tilde{P}^{\dot{a}}=\lambda_a K^{a\dot{a}}$, we find that the scalar bubble coefficient is given by:
\eq
c_s=-\frac{K^2}{2\pi i}\sum_i \frac{a_i\mathcal{F}(\lambda_e,\tilde{\lambda}_e)}{(n+1)!}\prod_{l=1}^{n}\left(\lambda_C^{\dot{a}}\frac{\partial}{\partial \tilde{P}^{\dot{a}}}\right)\int\frac{\AB{\lambda {\rm d}\lambda}\SB{\tilde{\lambda} {\rm d}\tilde{\lambda}}}{\SB{\tilde{P} \tilde{\lambda}}^2} \frac{\prod_j\AB{\lambda A_j}}{\prod_k\AB{\lambda B_k}}.
\eqe
The contour integration becomes
\eq
\begin{split}
&\prod_{l=1}^{n}\left(\lambda_C^{\dot{a}}\frac{\partial}{\partial \tilde{P}^{\dot{a}}}\right)\int\frac{\AB{\lambda {\rm d}\lambda}\SB{\tilde{\lambda} {\rm d}\tilde{\lambda}}}{\SB{\tilde{P} \tilde{\lambda}}^2}\frac{\prod_j\AB{\lambda A_j}}{\prod_k\AB{\lambda B_k}}\\&=\int\AB{\lambda {\rm d}\lambda}\left[-{\rm d}\tilde{\lambda}^{\dot{a}}\frac{\partial}{\partial\tilde{\lambda}^{\dot{a}}}\left(\prod_{l=1}^{n}\left(\lambda_C^{\dot{a}}\frac{\partial}{\partial \tilde{P}^{\dot{a}}}\right)\frac{\SB{\tilde{\lambda} \eta}}{\SB{\tilde{P} \tilde{\lambda}}\SB{\tilde{P} \eta}}\frac{\prod_j\AB{\lambda A_j}}{\prod_k\AB{\lambda B_k}}\right)\right]\\
&=-2\pi i\left.\sum_{|X\rangle=|B_k\rangle,K|\eta]}\left(\AB{\lambda X}\prod_{l=1}^{n}\left(\lambda_C^{\dot{a}}\frac{\partial}{\partial \tilde{P}^{\dot{a}}}\right)\frac{\SB{\tilde{\lambda} \eta}}{\ASB{\lambda | K | \tilde{\lambda}}\ASB{\lambda | K | \eta}}\frac{\prod_j\AB{\lambda A_j}}{\prod_k\AB{\lambda B_k}}\right)\right\vert_{|\lambda\rangle=| X\rangle},
\end{split}
\eqe
where $\eta$ is the reference spinor. In the last equation, we have used the equality
\eq
-{\rm d}\tilde{\lambda}^{\dot{a}}\frac{\partial}{\partial\tilde{\lambda}^{\dot{a}}}\frac{1}{\AB{\lambda B}}=2\pi\delta(\AB{\lambda B}),
\eqe
and the delta function is defined as:
\eq
\int\AB{\lambda {\rm d}\lambda}\delta(\AB{\lambda B})\mathcal{G}(\lambda)=-i\mathcal{G}(B)\,.
\eqe
Finally, we arrive at the general formula for the bubble coefficient:  
\eq
c_s\!=\!K^2\sum_i \frac{a_i\mathcal{F}(\lambda_e,\tilde{\lambda}_e)}{(n+1)!}\!\!\!\left.\sum_{|X\rangle=|B_k\rangle,K|\eta]}\!\!\!\left(\AB{\lambda X}\prod_{l=1}^{n}\left(\lambda_C^{\dot{a}}\tfrac{\partial}{\partial \tilde{P}^{\dot{a}}}\right)\!\frac{\SB{\tilde{\lambda} \eta}}{\ASB{\lambda | K | \tilde{\lambda}}\ASB{\lambda | K | \eta}}\frac{\prod_j\AB{\lambda A_j}}{\prod_k\AB{\lambda B_k}}\right)\right\vert_{|\lambda\rangle=| X\rangle} \! .
\eqe

\pagebreak

\section{Symmetrization}\label{app:cc}

We will derive a beautiful expression for $c_{ijkl}$ as a consequence of the generalized optical theorem in terms of the UV amplitudes, as described in \Sec{sec:generalizedunitarity}.
From \Eq{eq:cc}, the output of the dispersion relation is
\be 
c_{ijkl} + c_{ikjl} = \sum_m (m^{ij}m^{kl}+ m^{ik} m^{lj}),\label{eq:cccopy}
\ee
where the Wilson coefficients $c_{ijkl}$ satisfy the symmetries in \Eq{eq:sym}, $c_{ijkl}=c_{klij}=c_{jikl}=c_{ijlk}$, and we make no symmetry assumptions on the real matrices $m^{ij}$.
Subtracting the same identity with $i\leftrightarrow j$ exchanged, the result is
\be 
c_{ikjl}-c_{jkil}=\sum_{m}\left[m^{lk}(m^{ij}-m^{ji})+m^{ik}m^{lj}-m^{jk}m^{li}\right].\label{eq:ccsubtract}
\ee
On the other hand, simple relabeling allows us to write \Eq{eq:cccopy} as
\be
c_{iklj}+c_{ilkj}=\sum_{m}(m^{ik}m^{jl}+m^{il}m^{jk}).\label{eq:ccpermute}
\ee
Combining the two and using the symmetry properties in \Eq{eq:sym}, we find:
\be
c_{ijkl}=\sum_{m}\left(m^{ij}m^{(kl)}+m^{kj}m^{[il]}+m^{lj}m^{[ik]}\right).\label{eq:cc1} 
\ee
Simultaneously swapping $i\leftrightarrow k$ and $j\leftrightarrow l$, and using $c_{ijkl}=c_{klij}$, we have:
\be 
c_{ijkl}=\sum_{m}\left(m^{kl}m^{(ij)}+m^{il}m^{[kj]}-m^{jl}m^{[ik]}\right).\label{eq:cc2}
\ee
Adding Eqs.~\eqref{eq:cc1} and \eqref{eq:cc2}, we obtain:
\be 
c_{ijkl}=\sum_{m}\left(\frac{1}{2}m^{(ij)}m^{kl}+\frac{1}{2}m^{ij}m^{(kl)}+\frac{1}{2}m^{kj}m^{[il]}+\frac{1}{2}m^{il}m^{[kj]}+m^{[ik]}m^{[lj]}\right).\label{eq:cc3}
\ee
Taking Eq.~\eqref{eq:cc3} and simultaneously swapping $i\leftrightarrow l$ and $j\leftrightarrow k$, and using $c_{ijkl}=c_{lkji}$, the expression becomes:
\be 
c_{ijkl}=\sum_{m}\left(\frac{1}{2}m^{(kl)}m^{ji}+\frac{1}{2}m^{lk}m^{(ij)}-\frac{1}{2}m^{jk}m^{[il]}-\frac{1}{2}m^{li}m^{[kj]}+m^{[lj]}m^{[ik]}\right).\label{eq:cc4}
\ee
Finally, adding Eqs.~\eqref{eq:cc3} and \eqref{eq:cc4}, we find the elegant result of \Eq{eq:ccfinal}:
\be
c_{ijkl}=\sum_{m}\left(m^{(ij)}m^{(kl)}+m^{[il]}m^{[kj]}+m^{[ik]}m^{[lj]}\right). 
\ee

\pagebreak

\begin{figure}[t]
\begin{center}
\includegraphics[width=0.45\columnwidth]{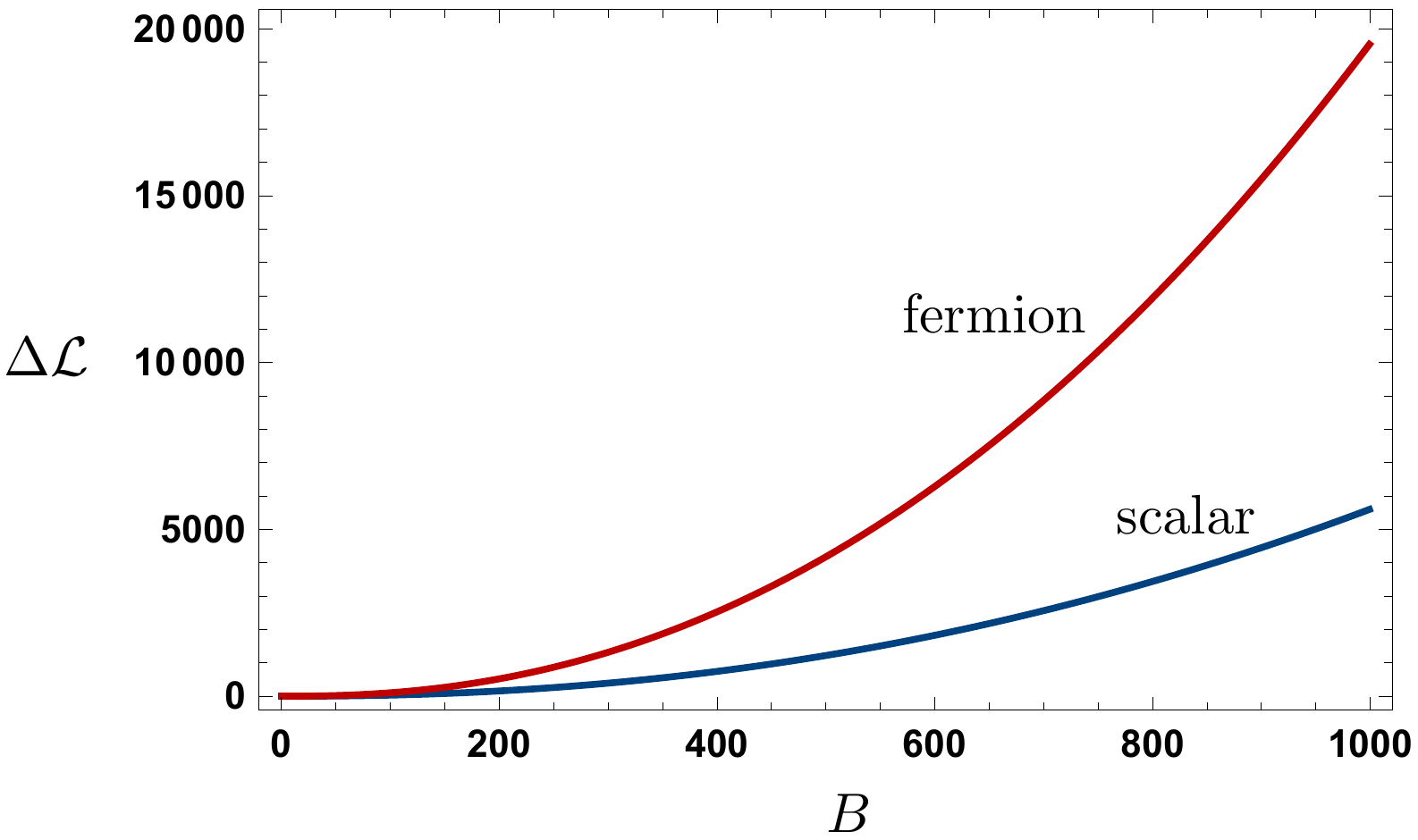}
\end{center}\vspace{-9mm}
\caption{Numerical integration of the Euler-Heisenberg Lagrangian in \Eq{eq:EulerHeisenberg} for large values of $|\vec{B}|$, with $m=e=1$. For both the fermion and scalar completion, $\Delta {\cal L}$ is positive and convex.
}
\label{fig:EulerHeisenberg}
\end{figure}

\section{Convexity of $\Delta {\cal L}$}\label{app:convex}
In addition to positivity, we further expect that $\Delta {\cal L}$ is a convex functional of the fields, as a consequence of causality.
Computing the equation of motion for a fluctuation of some field in the action, the first variation vanishes if we are expanding about a background that itself satisfies its equation of motion, so the second-order contribution from the fluctuation will dictate the dispersion relation.
Requiring subluminal propagation~\cite{Adams:2006sv} then enforces the second-derivative condition on $\Delta {\cal L}$.
In the case of a theory given by a polynomial $P$ in $(\partial\phi)^2$, $P'' \geq 0$ was shown to follow from causality in Ref.~\cite{Chandrasekaran:2018qmx}. This holds, for example, for the brane action ${\cal L} = -f^4 \sqrt{1+ (\partial\phi)^2}$.

A more pertinent example for the WGC is the Euler-Heisenberg Lagrangian. While the two leading $F^4$ terms are manifestly convex, let us consider the full one-loop action~\cite{Huet:2011kd}, resummed at arbitrarily high powers of $F_{\mu\nu}$. This can be written as a function of parameters $x,y$, where $x^2 - y^2 = B^2 - E^2$ and $xy=\vec{E}\cdot\vec{B}$, in terms of which ${\cal L} = -F^2/4 + \Delta {\cal L}$ where, depending on whether we integrate out a fermion or a scalar, we have, respectively,

\be
\begin{aligned}
\Delta {\cal L}_{\frac{1}{2}} &= \frac{ -1}{8\pi^2} \int_0^\infty \frac{{\rm d}s}{s^3}e^{-m^2 s} \left[{ \frac{(exs)(eys)}{\tanh(exs)\tan(eys)}- \frac{(es)^2 }{3}(x^2-y^2)- 1} \right]\\
\Delta {\cal L}_{0} &=  \frac{1}{16\pi^2} \int_0^\infty \frac{{\rm d}s}{s^3}e^{-m^2 s} \left[{ \frac{(exs)(eys)}{\sinh(exs)\sin(eys)}+ \frac{(es)^2}{6} (x^2-y^2)-1} \right].\label{eq:EulerHeisenberg}
\end{aligned}
\ee

For nonzero electric field, the integrand has singularities due to on-shell particle production, so one cannot resum all of the terms for large $F_{\mu\nu}$. If we set $E=0$ (i.e., only allow $x$ nonzero) in \Eq{eq:EulerHeisenberg}, the integrals can be computed for arbitrarily large magnetic field.
Doing so, we find that $\Delta {\cal L}$ is indeed positive and convex.
At small $B$, the integral reproduces the usual $F^4$ terms, while at large $B$, $\Delta{\cal L}\sim B^2$; see \Fig{fig:EulerHeisenberg}. Note that we can remain within the regime of validity of the EFT while considering large $B$, since we are taking a background of uniform $F_{\mu\nu}$.

\section{Perturbed extremal solutions}\label{GHSSol}
In this appendix, we derive the shift in the extremal condition of the magnetic GHS black hole given in \Eq{eq:Deltazdilaton}, for arbitrary dilaton coupling, in the Einstein-Maxwell-dilaton EFT of \Eq{eq:EMDEFT}.
We begin with the Einstein-Maxwell-dilaton Lagrangian in \Eq{eq:EMDL}.\footnote{Note that in \Eq{eq:EMDL}, the normalization for $F^2$ is $1/2$, rather than the usual $1/4$ in Maxwell theory. We will always choose this normalization when we consider Einstein-Maxwelld-dilaton theory, since it will help us avoid a square root in the extremal limit when $\lambda=1$.}
The low-energy EFT of the heterotic string has $\lambda=1$, while $\lambda=\sqrt{3}$ for KK compactification of Einstein gravity from five to four spacetime dimensions. Charged black holes will have nontrivial scalar profiles due to the scalar coupling to $F^2$.
For general $\lambda$, the magnetically-charged black hole solution was found by Garfinkle, Horowitz, and Strominger~\cite{GHS}: 
\eq
{\rm d}s^2= -e^{g(r)}{\rm d}t^2+e^{-g(r)}{\rm d}r^2+r^2f(r){\rm d}\Omega^2,
\eqe
where
\be 
\begin{aligned}
e^{g(r)}&=\left(1 - \frac{r_+}{r}\right)\left(1 - \frac{r_-}{r}\right)^{\frac{1 - \lambda^2}{1 + \lambda^2}}\\
f(r)&=\left(1 - \frac{r_-}{r}\right)^{\frac{2\lambda^2}{1 + \lambda^2}}.
\end{aligned}
\ee
The field strength is given by $F_{\mu\nu}{\rm d}x^\mu {\wedge} {\rm d}x^\nu =2 P \sin \theta \, {\rm d}\theta{\wedge} {\rm d}\varphi$ and the dilaton field profile is:
\eq
\phi=\phi_0 - \frac{\lambda}{1 + \lambda^2} \log\left(1 - \frac{r_-}{r}\right),
\eqe
where $\phi_0$ is the value of dilaton field at infinity, which will be set to zero for convenience. Here, $(r_-, r_+)$ are the inner and outer horizons, respectively,
\be\label{eq:DRPM}
\begin{aligned}
r_-&=\frac{1 + \lambda^2}{1 - \lambda^2}\left(M - \sqrt{M^2-\frac{1 - \lambda^2}{2}P^2}\right)\\
r_+&=M + \sqrt{M^2 - \frac{1 - \lambda^2}{2}P^2}.
\end{aligned}
\ee
For $\lambda\geq1$, which are the cases in which we are interested, the argument of the square root is always positive, and the singularity at $r=0$ is never naked. However, there is now also a singularity at $r=r_-$\,, as can be seen from the dilaton profile. Similarly, the Ricci scalar of the GHS solution also diverges at $r_-$: 
\eq
R_{\rm GHS}=\frac{2\lambda^2}{(1+\lambda^2)^2 r^4}\left(1-\frac{r_+}{r}\right)\left(1 - \frac{r_-}{r}\right)^{\frac{2}{1+\lambda^2} - 3}.
\eqe
Requiring the singularity at $r_-$ be behind the outer horizon, i.e., $r_-\leq r_+$, leads to the extremal condition $r_-=r_+$:
\eq
P=\sqrt{2(1+\lambda^2)}M.
\eqe 
Note that even if $\lambda<1$ where the square root in \Eq{eq:DRPM} can potentially be negative, the condition that $r_-\leq r_+$ still provides the most stringent bound on $P$. The electric solution can be obtained from the magnetic via duality transformations~\cite{Holzhey:1991bx}.

Let us now consider the effects of higher-derivative corrections, ${\cal L} = {\cal L}_{\rm EMD} + \Delta {\cal L}$. There are three classes of operators relevant to our consideration:
\be 
\begin{aligned}
\Delta {\cal L} =&\sum_{i=1}^2\,a_i e^{-6\lambda\phi}(F^4)_i + \sum_{i=1}^4 b_ie^{-4\lambda\phi}(\partial \phi^2F^2)_i + ce^{-2\lambda\phi}(\partial \phi)^4,
\end{aligned}
\ee
where $i$ label the independent tensor contractions,
\be 
\begin{aligned}
(F^4)_1&=(F_{\mu\nu}F^{\mu\nu})^2\\
(F^4)_2&=(F_{\mu\nu}\widetilde{F}^{\mu\nu})^2 \\
((\partial\phi)^2F^2)_1 &=\partial_{\mu}\phi\partial^{\mu}\phi F_{\nu\rho}F^{\nu\rho}\\((\partial\phi)^2F^2)_2&=\partial_{\mu}\phi\partial^{\nu}\phi F^{\mu\rho}F_{\nu\rho}\\((\partial\phi)^2F^2)_3&=\partial_{\mu}\phi\partial^{\mu}\phi F_{\nu\rho}\widetilde{F}^{\nu\rho}\\((\partial\phi)^2F^2)_4&=\partial_{\mu}\phi\partial^{\nu}\phi F^{\mu\rho}\widetilde{F}_{\nu\rho}.
\end{aligned}
\ee
Perturbing around a pure magnetic solution, only $(F_{\mu\nu}F^{\mu\nu})^2$, $\partial_{\mu}\phi\partial^{\mu}\phi F_{\nu\rho}F^{\nu\rho}$ and $(\partial \phi)^4$ modify the equations of motion. We deform the GHS solution by a linear perturbation and solve the corresponding Einstein equations up to first order in $a_1, b_1$ and $c$.

We begin with a spherically symmetric metric, transformed from string to Einstein frame,
\eq
{\rm d}s^2= e^{-2\lambda\phi(r)}\left[-e^{2\Phi(r)}{\rm d}t^2+e^{2\Lambda(r)}{\rm d}r^2+r^2{\rm d}\Omega^2\right],
\eqe
where
\be 
\begin{aligned}
\Phi(r)&=\frac{1}{2}\log\left[\left(1-\frac{r_+}{r}\right)\left(1-\frac{r_-}{r}\right)^{\frac{1-3\lambda^2}{1+\lambda^2}}\right] + \Phi_2(r)\\
\Lambda(r)&=-\frac{1}{2}\log\left[\left(1-\frac{r_+}{r}\right)\left(1-\frac{r_-}{r}\right)\right] + \Lambda_2(r)\\
\phi(r)&= - \frac{\lambda}{1+\lambda^2}\log\left(1-\frac{r_-}{r}\right) + \frac{1}{2\lambda}[\Phi_2(r)-\phi_2(r)].\\
\label{bgs}
\end{aligned}
\ee
Taking the background GHS solution to the extremal limit in which $r_-=r_+$ (which we will write as $r_0$), we find the following solution to the higher-derivative corrected Einstein, Maxwell, and dilaton equations of motion:
\be 
\begin{aligned}
\Phi_2(r)=&\,\frac{4 a_1 (\lambda ^2-1)}{15 (\lambda ^2+1)^3 r^4 (r-r_0)^3 r_0^2} \Big[-120\lambda^2 r^4 (r-r_0)^3 \log\left(1-\frac{r_0}{r}\right)-120 \lambda ^2 r^6 r_0  \\ &\qquad\qquad\qquad\qquad\qquad\qquad +300 \lambda ^2 r^5 r_0^2   -220 \lambda ^2 r^4 r_0^3+30 \lambda ^2 r^3 r_0^4+6 \lambda ^2 r^2 r_0^5
\\ &\qquad\qquad\qquad\qquad\qquad\qquad +(5 \lambda ^2+3) r r_0^6-3 r_0^7\Big]
\\&+\frac{b_1 \lambda ^2}{15 (\lambda ^2+1)^4 r^4 (r-r_0)^3 r_0^2} \Big[60(\lambda^2 - 3) r^4 (r-r_0)^3 \log\left(1-\frac{r_0}{r}\right) \\
& \qquad\qquad\qquad\qquad\qquad\qquad +5 (\lambda ^2-3) (12r^6 r_0-30 r^5 r_0^2 + 22 r^4 r_0^3 -3  r^3 r_0^4)\\& \qquad\qquad\qquad\qquad\qquad\qquad+3 (2 \lambda ^4+3 \lambda ^2+5) r^2 r_0^5  -(5 \lambda ^4+27 \lambda ^2+18) r r_0^6 
\\& \qquad\qquad\qquad\qquad\qquad\qquad + 6(2\lambda^2 +3) r_0^7\Big]
\\&+\frac{c \lambda ^4}{60 (\lambda ^2+1)^5 r^4 (r-r_0)^3 r_0^2} \Big[120 (\lambda ^4-3) r^4 (r-r_0)^3 \log \left(1-\frac{r_0}{r}\right)  
\\& \qquad\qquad\qquad\qquad\qquad\qquad +120 (\lambda ^4-3) r^6 r_0 -300 (\lambda ^4-3) r^5 r_0^2 \\
&\qquad\qquad\qquad\qquad\qquad\qquad+220 (\lambda ^4-3) r^4 r_0^3 -30 (\lambda ^4-3) r^3 r_0^4 
\\& \qquad\qquad\qquad\qquad\qquad\qquad +6 (\lambda ^4+4 \lambda ^2+5) r^2 r_0^5 -(15 \lambda ^4+52 \lambda ^2+33) r r_0^6 
\\& \qquad\qquad\qquad\qquad\qquad\qquad +3 (9 \lambda ^2+11) r_0^7\Big]
\\&+\frac{(\lambda ^2-1) r_0 \left(\sqrt{2} P_2-2 C_3 \sqrt{\lambda ^2+1}\right) \left[(2 \lambda ^2+3) r_0-3 (\lambda ^2+1) r\right]}{6 (\lambda ^2+1)^{3/2} (r-r_0)^3}
\\&  +\frac{C_4}{r_0-r}+C_2,
\end{aligned}
\ee

\vspace{1cm}

\be 
\begin{aligned}
\Lambda_2(r)=&\,\frac{4 a_1 r_0^4 \left[-(5 \lambda ^2+1) r_0+6 \lambda ^2 r+r\right]}{5 (\lambda ^2+1)^2 r^4 (r-r_0)^3} 
\\& +\frac{b_1 \lambda ^2 r_0^3 \left[-(14 \lambda ^2+19) r r_0+(5 \lambda ^2+9) r_0^2+10 (\lambda ^2+1) r^2\right]}{5 (\lambda ^2+1)^3 r^4 (r-r_0)^3}\\
&+\frac{c \lambda ^4 r_0^3 \left[-(34 \lambda ^2+39) r r_0+(15 \lambda ^2+19) r_0^2+20 (\lambda ^2+1) r^2\right]}{20 (\lambda ^2+1)^4 r^4 (r-r_0)^3} 
\\& +\frac{2 C_3 r (-r_0+\lambda ^2 r+r)+\sqrt{\frac{2}{1+\lambda^2}}P_2 r_0 \left[(\lambda ^2+1) r_0-(2 \lambda ^2+1) r\right]}{2 (r-r_0)^3},
\end{aligned}
\ee 
and
\be 
\begin{aligned}
\phi_2(r)=&\,a_1 \left[\frac{4 \lambda ^2 r (120 r^5-300 r_0 r^4+220 r_0^2 r^3-30 r_0^3 r^2-6 r_0^4 r-5 r_0^5)+12 (r_0-r) r_0^5}{15 (\lambda ^2+1)^2 r^4 (r-r_0)^3 r_0}\right. \\& \left. \qquad +\frac{32 \lambda ^2}{\left(\lambda ^2+1\right)^2 r_0^2}\log \left(1-\frac{r_0}{r}\right)\right]\\
&+b_1 \left\{\frac{\lambda ^2}{15 (\lambda ^2+1)^3 r^4 (r-r_0)^3 r_0} \left[-120 r^6+300 r_0 r^5-220 r_0^2 r^4+30 r_0^3 r^3 \right.\right. \\& \qquad\qquad\qquad\qquad\qquad\qquad\qquad \left. +\lambda ^2 (5 r_0-6 r) r_0^4 r+7 r_0^5 r+3 r_0^6\right]  \\& \qquad \left. -\frac{8 \lambda^2}{(\lambda ^2+1)^3 r_0^2} \log \left(1-\frac{r_0}{r}\right)\right\}\\
&+c \left\{\frac{\lambda ^4}{60 (\lambda ^2+1)^4 r^4 (r-r_0)^3 r_0}\left[(\lambda ^2{+}2) ({-}120 r^6{+}300 r_0 r^5{-}220 r_0^2 r^4{+}30 r_0^3 r^3)\right.\right. \\&\qquad\qquad\qquad\qquad\qquad\qquad\qquad  \left. -6 \lambda ^2 r_0^4 r^2+(15 \lambda ^2+17) r_0^5 r+3 r_0^6\right] \\
&\left.\qquad-\frac{2 \lambda ^4 \left(\lambda ^2+2\right)}{\left(\lambda ^2+1\right)^4 r_0^2}\log \left(1-\frac{r_0}{r}\right)\right\}
\\& -\frac{C_3 \left[3 (\lambda ^2+1) r^2-3 (\lambda ^2+1) r_0 r+\lambda ^2 r_0^2\right]}{3 (r-r_0)^3}
\\& +\frac{P_2 r_0 \left[3 (\lambda ^2+1) r-(2 \lambda ^2+3) r_0\right]}{3 \sqrt{2} \sqrt{\lambda ^2+1} (r-r_0)^3}+C_1.
\end{aligned}
\ee
Requiring the new solution to approach the Minkowski metric as $r\rightarrow \infty$ fixes $C_1= C_2=0$. Since the curvature in Einstein frame is singular on the extremal horizon, we will continue our analysis in the Jordan (string) frame,  which is related by a rescaling of the metric, $g^{\rm Einstein}_{\mu\nu}=e^{-2\lambda\phi}g^{\rm Jordan}_{\mu\nu}$.
Indeed, in the Jordan frame the Ricci scalar for the GHS solution yields
\eq
R^{\rm Jordan}_{\rm GHS} =\frac{8\lambda^2 r_0^2}{(1 + \lambda^2)^2r^4},
\eqe
which is regular on the extremal horizon $r=r_0$. Requiring this regularity be preserved for the deformed solution fixes the integral constants $C_3, C_4$ as
\be \label{intc}
\begin{aligned}
C_3= \frac{C_4}{1-\lambda^2} =  -\frac{16(1 + \lambda^2)^2a_1 + 4\lambda^2(1+\lambda^2)b_1 + \lambda^4c}{10\sqrt{2}(1 + \lambda^2)^{9/2}P} +\frac{P_2}{\sqrt{2 + 2\lambda^2}}.
\end{aligned}
\ee

To determine the new extremal solution, we note that for the undeformed case the operator $\nabla_{\alpha}R_{\mu\nu\rho\sigma}\nabla^{\alpha}R^{\mu\nu\rho\sigma}$  develops a double zero in the extremal limit,  
\eq
(\nabla R\nabla R)^{\rm Jordan}=(r-r_0)^2f(r)\,.
\eqe
We require that the perturbed extremal solution also have a double root. Expanding $\nabla_{\alpha}R_{\mu\nu\rho\sigma}\nabla^{\alpha}R^{\mu\nu\rho\sigma}$ as $r=r_0+\Delta r$ in the perturbed solution, up to quadratic order it is proportional to $(\Delta r)^2 + {\rm const.}$ (The linear term vanishes by construction.) Thus, in order for there to be a double root, the constant terms must vanish, which fixes $P_2$ to
\be 
\begin{aligned}
P_2  = \frac{16a_1(1 + \lambda^2)^2(1 + 3\lambda^2) - 4b_1\lambda^2(1 - \lambda^4) - c\lambda^4(3 + \lambda^2)}{2\sqrt{2}r_0(1 + \lambda^2)^{9/2}}.
\end{aligned}
\ee
We have now fixed all of the integral constants in the extremal limit of the perturbed solution in Jordan frame.

Identifying the mass from the $1/r$ term in $g_{tt}$ as $r\rightarrow\infty$, we have the mass and charge in perturbed black hole,
\eq
\begin{split}
&{\rm Mass}=M + \frac{1}{2}\left[(1 + \lambda^2)C_3 + C_4+\frac{2r_0(C_1+C_2)}{1 + \lambda^2}\right]\\
&{\rm Charge}=P+P_2,
\end{split}
\eqe
and the perturbed extremal bound given in \Eq{eq:Deltazdilaton},
\be\label{amcr}
\begin{aligned}
\frac{\rm Charge}{\sqrt{2(1+\lambda^2)}({\rm Mass})}= 1 + \frac{16(1 + \lambda^2)^2 a_1 + 4\lambda^2(1 + \lambda^2)b_1 + \lambda^4c}{10(1 + \lambda^2)^4P^2}.
\end{aligned}
\ee
This agrees with Eq.~(4.10) of Ref.~\cite{Loges:2019jzs}, with $h=2/(1+\lambda^2)$. 

\section{$\Delta \zeta \propto \Delta g^{rr}$ }\label{Proof}
In this appendix, we will demonstrate with explicit deformed black hole solutions that the extremality shift $\Delta \zeta$ is related to the metric shift $\Delta g^{rr}$ at fixed ADM charges via
\eq
\Delta \zeta =-\lim_{\zeta\rightarrow 1} \frac{\Delta g^{rr}}{\partial_\zeta \bar{g}^{rr}},\label{eq:DzetaDeltag}
\eqe
where $ \bar{g}^{rr}$ is the undeformed metric~\cite{Cheung:2018cwt,dyonic}.
Let us begin by writing the $rr$ component of the metric for the deformed extremal black hole as
\eq
g^{rr}=\bar{g}^{rr}+\delta g^{rr}.
\eqe
Writing the radial coordinate of the original horizon as $r_0$ and that of the perturbed horizon as $r_{\rm H}=r_{0}+\Delta r$, the new metric vanishes,
\be 
\begin{aligned}
g^{rr}(r_{\rm H})&=\bar{g}^{rr}(r_0) + \Delta r\,\partial_r\bar{g}^{rr}(r_0)  + \frac{1}{2}\Delta r^2\,\partial_r^2\bar{g}^{rr}(r_0)+\delta g^{rr}(r_{\rm H}).
\end{aligned}
\ee
On the extremal background metric, we have $\bar{g}^{rr}(r_0)=\partial_r\bar{g}^{rr}(r_0)=0$. Thus, we find:
\be 
\frac{1}{2}\Delta r^2\partial_r^2\bar{g}^{rr}(r_0)+\delta g^{rr}(r_0)=0.
\ee
Since the new horizon should be such that $\Delta r$ has a double root in the above equation, we conclude that  $\delta g^{rr}(r_0)=0$. We can rewrite $\delta g^{rr}$ as 
\be 
\delta g^{rr}=\Delta g^{rr}+\Delta \zeta \partial_\zeta \bar{g}^{rr},
\ee
where the $\partial_\zeta \bar{g}^{rr}$ is understood as parameterizing the undeformed metric in terms of the extremality parameter $\zeta=\sqrt{Q^2+P^2}/\sqrt{2}M$ for RN and $\zeta=P/\sqrt{2(1+\lambda^2)}M$ for the GHS black hole. Since we have concluded that $\delta g^{rr}=0$, we indeed find \Eq{eq:DzetaDeltag}.

Let us  now verify the above result for RN and GHS black holes. 
For RN, the leading-order correction to the metric is:
\be 
\begin{aligned}
\delta g^{rr}=&\,\frac{C}{2r} + \frac{P P_2+Q Q_2}{r^2} - a_1\frac{4(P^2-Q^2)^2}{5r^6}-a_2\frac{16P^2Q^2}{5r^6}.
\end{aligned}
\ee
We can identify $\partial_\zeta \bar{g}^{rr}$ as the second term in the above expression, $\Delta \zeta \partial_\zeta \bar{g}^{rr} =(2M^2/r^2)\Delta\zeta$, so we find $\Delta \zeta=(P P_2+Q Q_2)/2M^2$, and we will have $\delta g^{rr}(r_0) = 0$ when the integral constant is
\be 
\begin{aligned}
C=&-\frac{2\sqrt{2}(PP_2+QQ_2)}{(P^2+Q^2)^{1/2}} + \frac{32\sqrt{2}}{5(P^2+Q^2)^{5/2}}\left[a_1(P^2-Q^2)^2+4a_2P^2Q^2\right].
\end{aligned}
\ee
Moreover, as we are considering a fixed-mass black hole here, the correction $C$ to the ADM mass should vanish, and so the charges are fixed as
\eq
PP_2+QQ_2=\frac{16[a_1(P^2-Q^2)^2+4a_2P^2Q^2]}{5(P^2+Q^2)^2}\,.
\eqe
We then recover the required form of $\Delta \zeta$ in \Eq{eq:shiftRN}, thus verifying Eq.~\eqref{eq:DzetaDeltag}:
\eq
\Delta \zeta=\frac{16[a_1(P^2-Q^2)^2+4a_2P^2Q^2]}{5(P^2+Q^2)^3}.
\eqe 

For the GHS black hole, the leading-order correction to metric in Jordan frame is:
\be
\begin{aligned}
(\delta g^{rr})^{\rm Jordan}=&\,\frac{16 a \lambda^2 (\lambda^2+1)^2 +4 b \lambda^4(\lambda^2+1)+c \lambda^6 - 20 C_3 r_0(\lambda^2+1)^4}{10  (\lambda ^2+1)^4 r r_0}\\
&-\frac{16 a \lambda^2 (\lambda^2+1)^2+4 b \lambda^4(\lambda^2+1)+c \lambda^6+20 C_3 r_0\lambda^2 (\lambda^2+1)^4}{10  (\lambda ^2+1)^4  (r-r_0) r_0}\\
&-\frac{r_0^4\left[16 a  (\lambda^2 + 1)^2 (5\lambda^2 + 1)- 4 b  \lambda^2 (\lambda^2 + 1)(5\lambda^2 + 9) - c \lambda^4 (15\lambda^2 + 19)\right]}{10  (\lambda ^2+1)^4 r^6}\\
&+\frac{r_0^3\left[16 a \lambda^2 (\lambda^2 + 1)^2 -4 b \lambda^2(\lambda^2 + 1)(9\lambda^2 + 10)- c \lambda ^4 (19\lambda^2 + 20)\right]}{10  (\lambda ^2+1)^4 r^5}\\
&+\frac{(r^2 + r_0 r + r_0^2)\left[16 a \lambda^2 (\lambda^2 +1)^2+4 b \lambda^4 (\lambda^2 + 1)+c \lambda^6\right]}{10  (\lambda ^2+1)^4 r^4}
\\& +\frac{\sqrt{2} \sqrt{\lambda ^2+1} P_2 r_0}{r^2}+\frac{\sqrt{2} \lambda ^2 P_2}{\sqrt{\lambda ^2+1}  (r-r_0)}-\frac{\sqrt{2} \lambda ^2 P_2}{\sqrt{\lambda ^2+1} r}.
\end{aligned}
\ee
Since the equality $\delta g^{rr}(r_0) = 0$ should hold, the integral constant $C_3$ is not a free parameter anymore, but instead
\eq
C_3=\frac{P_2}{\sqrt{2} \sqrt{\lambda ^2+1}}-\frac{16 a \left(\lambda ^2+1\right)^2+4 b \left(\lambda ^4+\lambda ^2\right)+c \lambda ^4}{20 \left(\lambda ^2+1\right)^4 r_0}.
\eqe
The metric then simplifies:
\be 
\begin{aligned}
(\delta g^{rr})^{\rm Jordan}=&-\frac{r_0^4\left[16 a  (\lambda^2 + 1)^2 (5\lambda^2 + 1) - 4 b  \lambda^2 (\lambda^2 + 1)(5\lambda^2 + 9) -c \lambda^4 (15\lambda^2 + 19)\right]}{10 \left(\lambda ^2+1\right)^4 r^6}\\
&+\frac{r_0^3\left[16 a \lambda^2(\lambda^2 + 1)^2 -4b\lambda^2(\lambda^2 + 1)(9\lambda^2 + 10) -c \lambda^4(19\lambda^2 + 20) \right]}{10 \left(\lambda ^2+1\right)^4 r^5}
\\&+\frac{\left[r^3 + (r+r_0)(r^2 + r_0^2)\lambda^2\right]\left[16 a(\lambda^2 + 1)^2 +4 b \lambda^2(\lambda^2 + 1)+c \lambda^4\right]}{10 \left(\lambda ^2+1\right)^4 r^4 r_0}\\
&+\frac{\sqrt{2} \sqrt{\lambda ^2+1} P_2 r_0}{r^2}-\frac{\sqrt{2} \sqrt{\lambda ^2+1} P_2}{r}.
\end{aligned}
\ee
The last two terms can be identified with $\partial_\zeta(\bar{g}^{rr})^{\rm Jordan}$,
\eq
\Delta \zeta\partial_\zeta(\bar{g}^{rr})^{\rm Jordan}=\Delta \zeta \frac{\sqrt{2}r_0(r_0-r)}{r^2\sqrt{1+\lambda^2}},
\eqe
with $\Delta \zeta=P_2\sqrt{1+\lambda^2}/\sqrt{2}r_0$. Requiring that the correction to the ADM mass vanishes implies
\eq
(1+\lambda^2)C_3+C_4+\frac{2r_0(C_1+C_2)}{1+\lambda^2}=0,
\eqe
where the integral constants are already fixed as in Eq.~\eqref{intc}. The magnetic charge shift $P_2$ is then fixed as
\eq
P_2=\frac{16a(1+\lambda^2)^2+4b\lambda^2(1+\lambda^2)+c\lambda^4}{10P(1+\lambda^2)^4}.
\eqe
The extremal parameter $\zeta$ is shifted as
\be 
\Delta \zeta =1+\frac{16a(1+\lambda^2)^2+4b\lambda^2(1+\lambda^2)+c\lambda^4}{10P^2(1+\lambda^2)^4}\,,
\ee
matching \Eq{amcr}. This verifies \Eq{eq:DzetaDeltag}.

\pagebreak

\bibliographystyle{utphys-modified}
\bibliography{WGC}

\end{document}